\UseRawInputEncoding 
\documentclass[a4paper,fleqn,usenatbib]{mnras}

\usepackage[T1]{fontenc}
\usepackage{ae,aecompl}

\usepackage{graphicx}	
\usepackage{amsmath}	
\usepackage{amssymb}	
\usepackage{mathrsfs}                
\usepackage{booktabs}
\usepackage{threeparttable}
\usepackage{longtable}

\title[Luminosity and redshift distributions of GRBs]{Revisiting the luminosity and redshift distributions of long gamma-ray bursts}

\author[Lan et al.]{
Guang-Xuan Lan,$^{1,2}$
Jun-Jie Wei,$^{1,2}$\thanks{E-mail: jjwei@pmo.ac.cn (JJW)}
Hou-Dun Zeng,$^{1,3}$\thanks{E-mail: zhd@pmo.ac.cn (ZHD)}
Ye Li $^{1}$
and Xue-Feng Wu$^{1,2}$\thanks{E-mail: xfwu@pmo.ac.cn (XFW)}
\\
$^{1}$Purple Mountain Observatory, Chinese Academy of Sciences, Nanjing 210023, China\\
$^{2}$School of Astronomy and Space Science, University of Science and Technology of China, Hefei 230026, China\\
$^{3}$Key Laboratory of Dark Matter and Space Astronomy Purple Mountain Observatory, Chinese Academy of Sciences Nanjing 210023, China
}

\date{Accepted XXX. Received YYY; in original form ZZZ}

\pubyear{2020}

\begin{document}
\label{firstpage}
\pagerange{\pageref{firstpage}--\pageref{lastpage}}
\maketitle

\begin{abstract}
In this work, we update and enlarge the long gamma-ray burst (GRB) sample detected by the {\it Swift} satellite.
Given the incomplete sampling of the faint bursts and the low completeness in redshift measurement,
we carefully select a subsample of bright {\it Swift} bursts to revisit the GRB luminosity function (LF) and
redshift distribution by taking into account the probability of redshift measurement. Here we also explore
two general expressions for the GRB LF, i.e., a broken power-law LF and a triple power-law LF. Our results suggest
that a strong redshift evolution in luminosity (with an evolution index of $\delta=1.92^{+0.25}_{-0.37}$)
or in density ($\delta=1.26^{+0.33}_{-0.34}$) is required in order to well account for the observations,
independent of the assumed expression of the GRB LF. However, in a one-on-one
comparison using the Akaike information criterion, the best-fitting evolution model involving the triple power-law LF
is statistically preferred over the best-fitting one involving the broken power-law LF with a relative
probability of $\sim94.3$\% versus $\sim5.7$\%. Extrapolating our fitting results to the flux limit of
the whole {\it Swift} sample, and considering the trigger probability of {\it Swift}/Burst Alert Telescope in detail,
we find that the expectations from our evolution models provide a good representation
of the observed distributions of the whole sample without the need for any adjustment of the model free parameters.
This further confirms the reliability of our analysis results.
\end{abstract}




\section{Introduction}
\label{sec:Intro}
Gamma-ray bursts (GRBs) are the most luminous explosive transients in the cosmos, which can be detected from
our local universe up to extremely high redshifts \citep{2009Natur.461.1258S,2009Natur.461.1254T,2011ApJ...736....7C}.
Generally, long GRBs with durations $T_{\rm 90}>2$ s are believed to be powered by the core collapse of massive stars
(e.g., \citealt{1993AAS...182.5505W,1998ApJ...494L..45P,2006ARA&A..44..507W}), a perspective given strong support by
observations of several long GRBs associated with supernovae (e.g., \citealt{2003Natur.423..847H,2003ApJ...591L..17S}).
The collapsar model suggests that the births of long GRBs should trace the cosmic star formation history (e.g.,
\citealt{1997ApJ...486L..71T,2004RvMP...76.1143P,2004IJMPA..19.2385Z,2007ChJAA...7....1Z}), and long GRBs may therefore be
excellent tools to probe the star formation rate (SFR) at high redshifts ($z\geq4$; e.g., \citealt{2007ApJ...671..272C,
2008ApJ...683L...5Y,2009ApJ...705L.104K,2009MNRAS.400L..10W,2013A&A...556A..90W,2014MNRAS.439.3329W}).

In the past few years, the properties of the population of long GRBs and their evolution with cosmic time have been
widely investigated by different authors (e.g., \citealt{2001ApJ...548..522P,2002ApJ...574..554L,2019MNRAS.488.5823L,2004ApJ...611.1033F,2005ApJ...619..412G,
2005MNRAS.364L...8N,2006MNRAS.372.1034D,2007JCAP...07..003G,2007ApJ...661..394L,2007ApJ...656L..49S,
2008ApJ...673L.119K,2009MNRAS.396..299S,2012ApJ...749...68S,2010ApJ...711..495B,2010MNRAS.407.1972C,
2010MNRAS.406..558Q,2010MNRAS.406.1944W,2011MNRAS.416.2174C,2011MNRAS.417.3025V,2012ApJ...745..168L,
2012ApJ...744...95R,2013ApJ...772L...8T,2014MNRAS.439.3329W,2015ApJ...806...44P,2015MNRAS.454.1785T,
2015ApJS..218...13Y,2016ApJ...820...66D,2016A&A...587A..40P,2017IJMPD..2630002W,2018MNRAS.473.3385P,
2019MNRAS.488.4607L,2020MNRAS.493.1479L,2021arXiv210316347D,2021A&A...649A.166P}).
There is a general agreement on the fact that GRBs may have experienced some kind of evolution with redshift,
whereas the nature and the level of such evolution are uncertain and still a matter of debate. It should be emphasized
that the evolution features of GRBs are inferred from the statistics of the observational samples. However,
as is well known, the observational samples inevitably subject to various instrumental selection effects and
observational biases \citep{2003AJ....125.2865B}. It is difficult to reveal the intrinsic distributions of
long GRBs and their evolution features from these samples. For example, a faint GRB with peak flux slightly
over the instrumental detection threshold would not always make a trigger. Incomplete sampling of the faint
bursts is thus a ubiquitous problem in observations. In addition, although the number of measured GRB redshifts
has increased considerably in the {\it Swift} era, the bursts with redshift determinations account for only $\sim1/3$
of all GRBs detected by {\it Swift}/Burst Alert Telescope (BAT). The low completeness level in redshift measurement
would lead to biases in shaping GRB redshift distribution \citep{2007A&A...470..515F,2012ApJ...749...68S}.
Therefore, the distributions of long GRBs through cosmic time can only be unbiasedly studied by using a complete sample that
is capable of adequately representing this population \citep{2012ApJ...749...68S,2016A&A...587A..40P,2019MNRAS.488.4607L},
or by using an incomplete sample but carefully taking into account various instrumental selection effects.

A few attempts have been made to construct a complete flux-limited sample of long GRBs. For instance,
\cite{2012ApJ...749...68S} defined a well-selected subsample of bright {\it Swift} GRBs, which is composed
of 58 bursts with favorable observing conditions for redshift measurement as proposed in \cite{2006A&A...447..897J}
and with 1-s peak photon flux $P\geq2.6$ photons cm$^{-2}$ s$^{-1}$. This sample, which is complete in
peak flux by definition, after careful selection turned out to be also highly complete ($\sim90$\%) in
redshift (i.e., 52 out of 58 bursts have measured $z$). \cite{2016A&A...587A..40P} extended the complete
sample of \cite{2012ApJ...749...68S} with additional bursts that satisfy the same selection criteria.
This extended sample contains 99 bursts, 81 of them with measured redshift and luminosity for a
completeness level of $\sim82$\%. Though the complete sample can provide a solid basis for the statistical
study of the long-GRB population \citep{2012ApJ...749...68S,2016A&A...587A..40P,2019MNRAS.488.4607L},
its current sample size is admittedly small. We have to acknowledge the statistical limitations of
the study due to the small sample size. On the other hand, many previous works relied on the use of
incomplete samples. Despite they have relatively large sample sizes (hundreds of bursts), the instrumental
selection effects and possible observational biases have not been handled very well \citep{2021MNRAS.504.4192B}.

In this work, we update and enlarge the GRB sample observed by {\it Swift}/BAT. Our principal goal in this paper
is to make use of the latest incomplete sample to revisit the GRB luminosity function (LF) and redshift
distribution by considering various instrumental selection effects, including the incomplete sampling of
the faint bursts and the probability of redshift measurement. In order to examine our analysis method's
effectiveness and reliability, we also extrapolate our findings to the detection limit of {\it Swift} by
considering the trigger probability of {\it Swift}/BAT, and the result shows that our method is effective and reliable.

This paper is arranged as follows. In Section~\ref{sec:sample}, we assemble our {\it Swift} GRB sample and estimate
the probability of redshift measurement for a GRB. Our analysis method is described in Section~\ref{sec:method},
and the model results are presented in \ref{sec:results}. In Section~\ref{sec:trigger}, we extrapolate our fitting
results to the flux limit of the whole redshift sample with the trigger probability of {\it Swift}/BAT. Our conclusions
are drawn in Section~\ref{sec:summary}. Throughout this paper a flat $\Lambda$CDM cosmological model with
$H_{0}=70$ km s$^{-1}$ Mpc$^{-1}$, $\Omega_{\rm m}=0.3$, and $\Omega_{\Lambda}=0.7$ is adopted.

\section{{\it Swift} GRB Sample and Redshift Measurement Probability}
\label{sec:sample}

\subsection{The sample}
\label{subsec:sample}
\begin{figure}
\vskip-0.1in
  \centering
  \includegraphics[angle=0,scale=0.6]{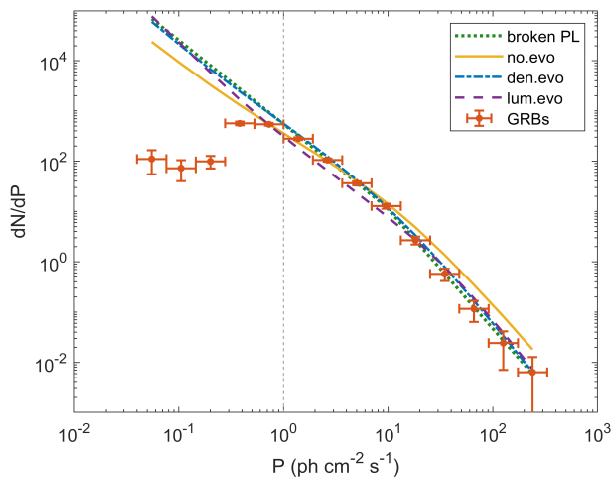}\\
  \caption{The peak-flux distribution of 1111 long GRBs detected by {\it Swift}/BAT. The vertical dotted line
  represents the limiting flux (1 photons cm$^{-2}$ s$^{-1}$) above which the incomplete sampling of the
  faint bursts induced by the instrumental selection effect becomes negligible.
  The green dotted line shows the best-fitting result to the observed peak-flux distribution with
  $P\geq1$ photons cm$^{-2}$ s$^{-1}$ by using a broken power-law function.
  Other curves show the expected flux distributions from different best-fitting models with the assumed tripe
  power-law LF: no evolution model (orange solid line), density evolution model (blue dot-dashed line),
and luminosity evolution model (purple dashed line).}
  \label{fig1}
\end{figure}

Since the launch of the {\it Swift} satellite on 2004 November \citep{2004ApJ...611.1005G}, more than one thousand GRBs have
been detected. Our sample is only limited to those long GRBs detected by {\it Swift}. This is because it is a homogeneous sample
whose size is large enough to do a reliable statistical analysis. We collect long GRBs with durations $T_{90}>2$ s up to
2019 November. Moreover, we require 1-s peak photon flux for analysis and it turns out to be 1111 GRBs. Figure~\ref{fig1}
shows the peak-flux distribution of all the 1111 bursts, ranging from 0.04 to 331 photons cm$^{-2}$ s$^{-1}$. Of these 1111,
429 bursts have accurate redshift measurements. The redshift completeness level is only $\sim39\%$. We calculate the peak
bolometric luminosities in the 1--$10^{4}$ keV rest-frame energy range for those GRBs having redshifts
(see specific formula in \citealt{2015ApJ...812...33S}). Within long GRBs, the low-luminosity GRBs (LLGRBs;
$L<10^{49}$ erg $\rm s^{-1}$) have been claimed to be a distinct population \citep{2004Natur.430..648S,2006ApJ...645L.113C,2007MNRAS.382L..21C,2007ApJ...662.1111L}.
There are five bursts with $L<10^{49}$ erg $\rm s^{-1}$ in our redshift sample. The luminosity distribution
of our sample shows that the LLGRBs are not the straightforward extension of high-luminosity GRBs (HLGRBs) to low luminosities.
Compared to the first component of HLGRBs, the incline of LLGRBs is slightly steeper and their normalization is a little
bit lower. Given the slight mismatch between LLGRBs and HLGRBs, we discard these five LLGRBs.
We have verified that our results are only weakly sensitive to the abandon of five LLGRBs.
The remaining 424 long GRBs with known redshifts are presented in Table~\ref{tabA1}
which includes the following information for each GRB: its name, the peak photon flux $P$ in the observer frame {\it Swift}/BAT
[15--150] keV energy band, the redshift $z$, the spectral parameters (low- and high-energy photon spectral indices $\alpha$ and
$\beta$, and the observed peak energy $E_p$), and the peak bolometric luminosity $L$.
The spectrum is a cut-off power law if only the low energy photon index $\alpha$ and the peak energy $E_p$ are reported,
and a Band function \citep{1993ApJ...413..281B} if the high energy photon index $\beta$ is also given.
For those GRBs without spectral information, we adopt the typical values of the Band spectrum ($\alpha=-1$, $\beta=-2.3$,
and $E_p=250$ keV; \citealt{2000ApJS..126...19P,2006ApJS..166..298K}) to calculate $L$.

The {\it Swift}/BAT trigger is complicated and its sensitivity for GRBs is difficult to parametrize exactly
\citep{2006ApJ...644..378B}. In practice, not all faint bursts with peak flux slightly over the instrument threshold
would be triggered successfully. As shown in Figure~\ref{fig1}, due to incomplete sampling of the faint bursts,
the peak-flux distribution is significantly deviated from an ideal power law at the low end. Thus, given this
incomplete sampling, the intrinsic peak-flux distribution is necessarily different from the observed one.
A simple approach to recovering the intrinsic peak-flux distribution is obtained by studying the sample above
$P\approx 1$ photons cm$^{-2}$ s$^{-1}$. Indeed, above this flux limit, the instrumental selection effect affecting
the incomplete sampling is negligible (see Figure~\ref{fig1}). Thus, we define a sharp trigger threshold:
the trigger probability $\theta_{\gamma}$ of bursts with $P\geq1$ photons cm$^{-2}$ s$^{-1}$ is 100\%. Above the
flux limit ($P_{\rm lim}=1$ photons cm$^{-2}$ s$^{-1}$), BAT detects 733 GRBs. 302 of them have measured redshift
so that the redshift completeness level is slightly increased to $\sim41\%$. Still, this level of completeness
is far too low to permit reliable population studies. Thus, the sample of the 302 GRBs with $P\geq1$ photons
cm$^{-2}$ s$^{-1}$ will be used to construct the GRB LF and its evolution with cosmic time by considering
the probability for redshift measurement.

\subsection{Redshift measurement probability}
\begin{figure}
\vskip-0.1in
  \centering
  \includegraphics[angle=0,scale=0.6]{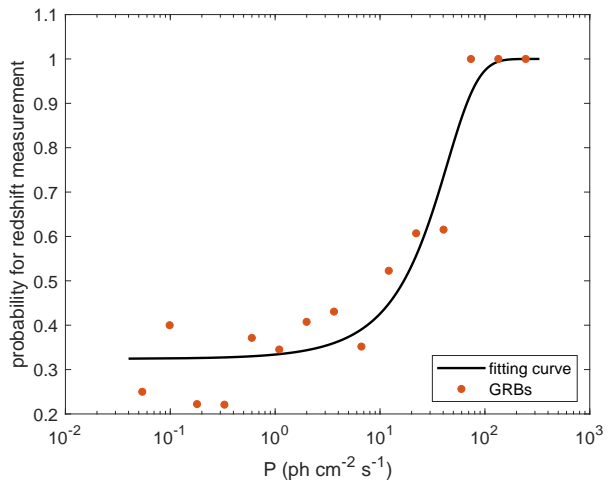}\\
  \caption{Redshift measurement probability as a function of the peak photon flux for {\it Swift} long GRBs.
  The best-fitting curve is parametrized in equation~(\ref{eq:zP}).}
  \label{fig2}
\end{figure}

The low completeness in redshift may disguise the intrinsic redshift distribution of GRBs
\citep{2007A&A...470..515F,2012ApJ...749...68S}.
In order to mitigate the incompleteness problem, we empirically model probability of redshift measurement
for a {\it Swift} detected GRB. Actually, the redshift measurement is quite complex, depending on
several artificial effects, such as the optical follow-up observation, spectral line confirmation,
and so on \citep{2003AJ....125.2865B}. It is difficult to model favorable observing conditions for
the optical follow-up searches. But \cite{2006ApJ...644..378B} suggested that the probability to
measure a redshift could still be empirically modeled, and it should be a function of the burst flux,
duration, and other factors. We restrict ourselves here only to the dependence of the probability on
the peak flux, since this is the observed quantity we adopt for the analysis in this paper.
Indeed, observations suggest redshifts are preferentially measured for brighter GRBs.
Figure~\ref{fig2} shows the redshift measurement probability $\theta_z$ as a function of the peak flux $P$,
in which the probabilities are estimated by the fraction of the number of bursts with redshift measurement in each bin to
the overall number of detected bursts in the corresponding bin. It is clear that the probability of redshift
determination for the burst in our sample is mildly relate to $P$, reflecting the observational fact that
the probability increases with increasing $P$ values. Here we use an empirical function to model
the redshift measurement probability
\begin{equation}
\theta_z(P)=\frac{1}{1+\xi_z\kappa_z^P}\;,
\label{eq:zP}
\end{equation}
where the best-fitting parameters are $\xi_z= 2.09\pm0.26$ and $\kappa_z=0.96\pm0.01$ at 68\% confidence level.

\section{Analysis Method}
\label{sec:method}

A maximum likelihood method firstly introduced by \cite{1983ApJ...269...35M} is adopted in our analysis.
The model free parameters can be optimized by maximizing the likelihood function (e.g.,
\citealt{1998ApJ...496..752C,2006ApJ...643...81N,2009ApJ...699..603A,2012ApJ...751..108A,2010ApJ...720..435A,
2014MNRAS.441.1760Z, 2016MNRAS.462.3094Z,2019MNRAS.488.4607L,2019MNRAS.490..758Q}), i.e.,
\begin{equation}
\mathcal{L}=\exp(-N_{\rm exp})\prod ^{N_{\rm obs}}_{i=1} \Phi(L_i,z_i,t_i)\;,
\label{eq:Likelihood}
\end{equation}
where $N_{\rm exp}$ denotes the expected number of GRB detections, $N_{\rm obs}$ represents the observed number
of the sample, and $\Phi(L,z,t)$ is the observed rate of bursts per unit time at redshift $z\sim z+dz$ with luminosity
$L\sim  L+dL$, which is given by
\begin{eqnarray}
\Phi\left(L,z,t\right)&=& \nonumber\frac{{\rm d}^3 N}{{\rm d}t{\rm d}z{\rm d}L}=\frac{{\rm d}^3 N}{{\rm d}t{\rm d}V{\rm d}L}\times\frac{{\rm d}V}{{\rm d}z}\\
&=& \frac{\Delta \Omega}{4\pi}\theta(P)\frac{\psi(z)}{1+z}\phi(L,z)\times\frac{{\rm d}V}{{\rm d}z}\;,
\label{eq:DL_age}
\end{eqnarray}
where $\Delta \Omega=1.33$ sr is the {\it Swift}/BAT field of view, $\theta(P)\equiv \theta_{\gamma}(P)\theta_{z}(P)$
is the detection efficiency which represents the probability to trigger and measure redshift for a burst with
peak flux $P$, $\psi(z)$ is the comoving event rate of GRBs (in units of $\rm Mpc^{-3}$ $\rm yr^{-1}$) as
a function of $z$, $(1+z)^{-1}$ reflects the cosmological time dilation, $\phi(L,z)$ is the normalized GRB LF which
whether evolves with redshift dependents on the assumed model (see more below),
and ${\rm d}V(z)/{\rm d}z=4\pi c D_L^2(z)/[H_{0}(1+z)^2\sqrt{\Omega_{\rm m}(1+z)^3+\Omega_{\Lambda}}]$
is the comoving volume element in a flat $\Lambda$CDM model. The $D_L(z)$ is the luminosity distance at $z$.

Within the context of the collapsar origin, the formation of each GRB just means the death of a short-lived massive star.
Thus, the GRB formation rate $\psi(z)$ should in principle be connected to the cosmic SFR $\psi_\star(z)$, i.e.,
\begin{equation}
\psi(z)=\eta \psi_\star(z)\;,
\end{equation}
where $\eta$ is the GRB formation efficiency that accounts for the number of GRBs formed per solar mass in stars
and has units of ${\rm M}_{\odot}^{-1}$. The SFR $\psi_\star(z)$ (in units of ${\rm M}_{\odot}$ $\rm yr^{-1}$ $\rm Mpc^{-3}$)
can be expressed approximately by \citep{2006ApJ...651..142H,2008MNRAS.388.1487L}
\begin{equation}
\psi_{\star}(z)=\frac{0.0157+0.118z}{1+(z/3.23)^{4.66}}\;.
\end{equation}

For the GRB LF $\phi (L,z)$, we use the general expression in term of a broken power law:
\begin{equation}
 \phi (L,z)  = \frac{A}{\ln(10)L}\left\{\begin{array}{l}
{\left(\frac{{{L}}}{{{L_{c}(z)}}}\right)^a};\,\,{L} \le {L_{c}(z)}\;,\\
{\left(\frac{{{L}}}{{{L_{c}(z)}}}\right)^b};\,\,{L} > {L_{c}(z)}\;,
\end{array} \right.
\end{equation}
where $A$ is a normalization constant, and $a$ and $b$ are the power-law indices before and after the break luminosity
$L_c$. If the broken power-law model does not reproduce the data well, we introduce a triple power-law form, i.e.,
\begin{equation}
\phi(L,z)=\frac{A}{\ln (10) L}\left\{\begin{array}{l}
\left(\frac{L}{L_{c_{1}}(z)}\right)^{a};\,\,\,\,\,\,\,\,\,\,\,\,\,\,\,\,\,\,\,\,\,\,\,\,\,\,\, L\leq L_{c_{1}}(z)\;, \\
\left(\frac{L}{L_{c_{1}}(z)}\right)^{b};\,\,\,\,\,\,\,\,\,\,\,\,\,\,\,\,\,\,\,\,\,\,\,\,\,\,\, L_{c_{1}}(z)< L \leq L_{c_{2}}(z)\;, \\
\left(\frac{L_{c_{2}}(z)}{L_{c_{1}}(z)}\right)^{b}\left(\frac{L}{L_{c_{2}}(z)}\right)^{c};\, L>L_{c_{2}}(z)\;,
\end{array}\right.
\end{equation}
where $a$, $b$, and $c$ are the power-law indices for three segments, and $L_{c_1}$ and $L_{c_2}$ are
the two break luminosities.

Considering the flux threshold that used to define 100-percent trigger (i.e., $P_{\rm lim}=1$ photons cm$^{-2}$ s$^{-1}$
in the 15--150 keV energy band), the expected number of GRBs can be expressed as
\begin{equation}
\begin{aligned}
N_{\rm exp} = \frac{\Delta \Omega T}{4\pi} \int ^{z_{\rm max}} _{0} & \int ^{L_{\rm max}}_{{\rm max}[L_{\rm min},L_{\rm lim}(z)]} \theta\left(P(L,z)\right)\frac{\psi(z)}{1+z} \\
&\times\phi(L,z){\rm d}L{\rm d}V(z)\;.
\end{aligned}
\label{eq:Nexp}
\end{equation}
where $T\sim15$ yr is the time of activity of {\it Swift} that covers our sample.
Since $z < 10$ for the current BAT sample, the maximum redshift for our analysis is $z_{\rm max}=10$.
The luminosity function is assumed to extend from $L_{\rm min}=10^{49}$ erg $\rm s^{-1}$ to $L_{\rm max}=10^{55}$
erg $\rm s^{-1}$ \citep{2015MNRAS.447.1911P}. The luminosity threshold appearing in Equation~(\ref{eq:Nexp})
can be estimated through
\begin{equation}
L_{\rm lim}(z)=4\pi D_L^2(z) P_{\rm lim} \frac{\int^{10^4/(1+z)\;{\rm keV}}_{1/(1+z)\;{\rm keV}} EN(E){\rm d}E}{\int^{150\;{\rm keV}}_{15\;{\rm keV}} N(E){\rm d}E}\;,
\end{equation}
where $N(E)$ is the GRB photon spectrum. We adopt a typical Band function spectrum with low- and high-energy
spectral indices $-1$ and $-2.3$, respectively \citep{1993ApJ...413..281B,2000ApJS..126...19P,2006ApJS..166..298K}.
In order to broadly estimate the spectral peak energy $E_{p}$ for a given $L$, we assume the validity of
the empirical $E_{p}$--$L$ correlation \citep{2004ApJ...609..935Y,2012MNRAS.421.1256N}, i.e.,
$\log \left[E_{p}(1+z)\right]=-25.33+0.53\log L$.

Due to the fact that (i) the connection between the GRB formation rate and the cosmic SFR may not be unbiased
and (ii) the GRB LF may evolve with redshift, we introduce an extra evolving factor $(1+z)^{\delta}$ in our above
equations, where $\delta$ is an (unknown) evolution parameter. We consider three different models.
In the first model, the GRB formation rate purely follows the SFR, i.e., $\psi(z)=\eta \psi_\star(z)$,
and the LF does not evolve with redshift, i.e., $L_{c_{i}}(z)=L_{c_{i},0}={\rm constant}$.
In the second one, while the GRB formation rate is still proportional to the SFR, the break luminosity in the
GRB LF increases with redshift as $L_{c_{i}}(z)=L_{c_{i},0}(1+z)^{\delta}$. Finally, we consider a model in which the
formation rate follows the SFR in conjunction with an extra evolving factor, i.e., $\psi(z)=\eta \psi_\star(z)(1+z)^{\delta}$,
and the LF is taken to be non-evolving.

For a given model, the free parameters can be optimized by maximizing the likelihood function (Equation~(\ref{eq:Likelihood})).
Since there are multiple parameters in our various models, the Markov Chain Monte Carlo (MCMC) sampling technology
is employed to derive the best-fitting values and the associated $1\sigma$ uncertainties for the model parameters.
We adopt the MCMC code from GWMCMC,\footnote{https://github.com/grinsted/gwmcmc} which is an implementation of
the Goodman and Weare 2010 Affine invariant ensemble MCMC sampler \citep{2010CAMCS...5...65G,2013PASP..125..306F}.

\section{Results}
\label{sec:results}

Using the above analysis method, we can optimize the values of the model free parameters,
including the GRB LF, the evolution parameter, and the GRB formation efficiency $\eta$. The best-fitting parameters
and their corresponding $1\sigma$ confidence level for different models are given in Table~\ref{tab1}. To determine
statistically which of the models is preferred by the observational data, we also report the log-likelihood value
$\ln\mathcal{L}$ and the Akaike Information Criterion (AIC) score in the last two columns of Table~\ref{tab1}.
For each fitted model, the AIC score is evaluated as ${\rm AIC}=-2\ln\mathcal{L}+2n$, where $n$ is the number of
free parameters \citep{1974ITAC...19..716A,2007MNRAS.377L..74L}. With ${\rm AIC}_i$ characterizing model $\mathcal{M}_i$,
the unnormalized confidence that this model is correct is the Akaike weight $\exp(-{\rm AIC}_{i}/2)$.
Informally, $\mathcal{M}_i$ has the relative probability
\begin{equation}
P(\mathcal{M}_i)=\frac{\exp(-{\rm AIC}_{i}/2)}{\exp(-{\rm AIC}_{1}/2)+\exp(-{\rm AIC}_{2}/2)}
\end{equation}
of being the true model in a one-on-one comparison. Thus, the difference ${\rm AIC}_{2}-{\rm AIC}_{1}$ determines
the extent to which $\mathcal{M}_1$ is preferred over $\mathcal{M}_2$.

\newcommand{\tabincell}[2]{\begin{tabular}{@{}#1@{}}#2\end{tabular}}
\begin{table*}
\centering
\caption{Best-fitting parameters for different models.}
\label{tab1}
\resizebox{\textwidth}{!}{
\begin{tabular}{|l|c|c|c|c|c|c|c|c|}
\hline
  Model &  Evolution parameter & $\eta$ & $a$ & $b$ & $c$ & $\log{L_c}$ & $\ln\mathcal{L}$ & AIC \\
        &                      & ($10^{-8}$ ${\rm M}_{\odot}^{-1}$) &  &  &  & (erg $\rm s^{-1}$) &  &  \\
\hline
 \multicolumn{9}{|c|}{Broken power-law luminosity function}\\
\hline
  No evolution&  $\cdot\cdot\cdot$ & $ 5.21^{+0.91}_{-0.95}  $ &$-0.38^{+0.04}_{-0.04} $ & $ -1.63^{+0.26}_{-0.26}   $
  & $\cdot\cdot\cdot$ &$ 52.90^{+0.10}_{-0.10} $ & -70.15 & 148.30 \\
  Luminosity evolution &  $\delta=1.85^{+0.24}_{-0.23}  $ & $ 3.58^{+0.73}_{-0.80} $ &$ -0.46^{+0.08}_{-0.08}  $& $ -1.30^{+0.18}_{-0.15} $ & $\cdot\cdot\cdot$  & $51.62^{+0.19}_{-0.20}   $& -42.08 & 94.16\\
  Density evolution &  $\delta=1.43^{+0.22}_{-0.20}  $ & $3.79^{+0.80}_{-0.75}   $ &$-0.60^{+0.05}_{-0.05}   $ & $ -1.65^{+0.27}_{-0.28}$ & $\cdot\cdot\cdot$  & $52.98^{+0.11}_{-0.12} $& -43.63  & 97.26\\
  \hline
     \multicolumn{9}{|c|}{Triple power-law luminosity function}\\
  \hline
  No evolution&  $\cdot\cdot\cdot$ & $15.45^{+7.79}_{-7.65}   $ &$-1.62^{+0.73}_{-0.69}   $ & $-0.35^{+0.04}_{-0.04}  $ &$ -1.65^{+0.24}_{-0.26}   $ &$ 49.99^{+0.35}_{-0.42} $, $ 52.92^{+0.09}_{-0.10}  $ & -67.06  & 146.12 \\
  Luminosity evolution &  $\delta= 1.92^{+0.25}_{-0.37}  $ & $ 9.59^{+4.71}_{-4.15}   $ &$-1.40^{+0.55}_{-0.62} $ & $ -0.25^{+0.12}_{-0.13} $ &$-1.17^{+0.08}_{-0.29}  $  &$50.16^{+0.37}_{-0.42} $, $ 51.61^{+0.15}_{-0.29} $ &-37.27 & 88.54 \\
  Density evolution &  $\delta=1.26^{+0.33}_{-0.34}    $ & $8.50^{+3.65}_{-3.68}$ &$ -0.97^{+0.19}_{-0.24} $ & $-0.49^{+0.09}_{-0.08}$ & $-1.64^{+0.26}_{-0.25} $  &$ 50.55^{+0.20}_{-0.21} $, $ 52.92^{+0.11}_{-0.11} $ & -40.68& 95.36 \\
  \hline
\end{tabular}
}
\begin{description}
  \item[\emph{Note.}] {Parameter values were computed as the medians of the best-fitting parameters to
  the Monte Carlo sample. Errors show the 68\% containment regions around the median values}
\end{description}
\end{table*}

\begin{figure*}
\vskip-0.1in
\includegraphics[angle=0,scale=0.55]{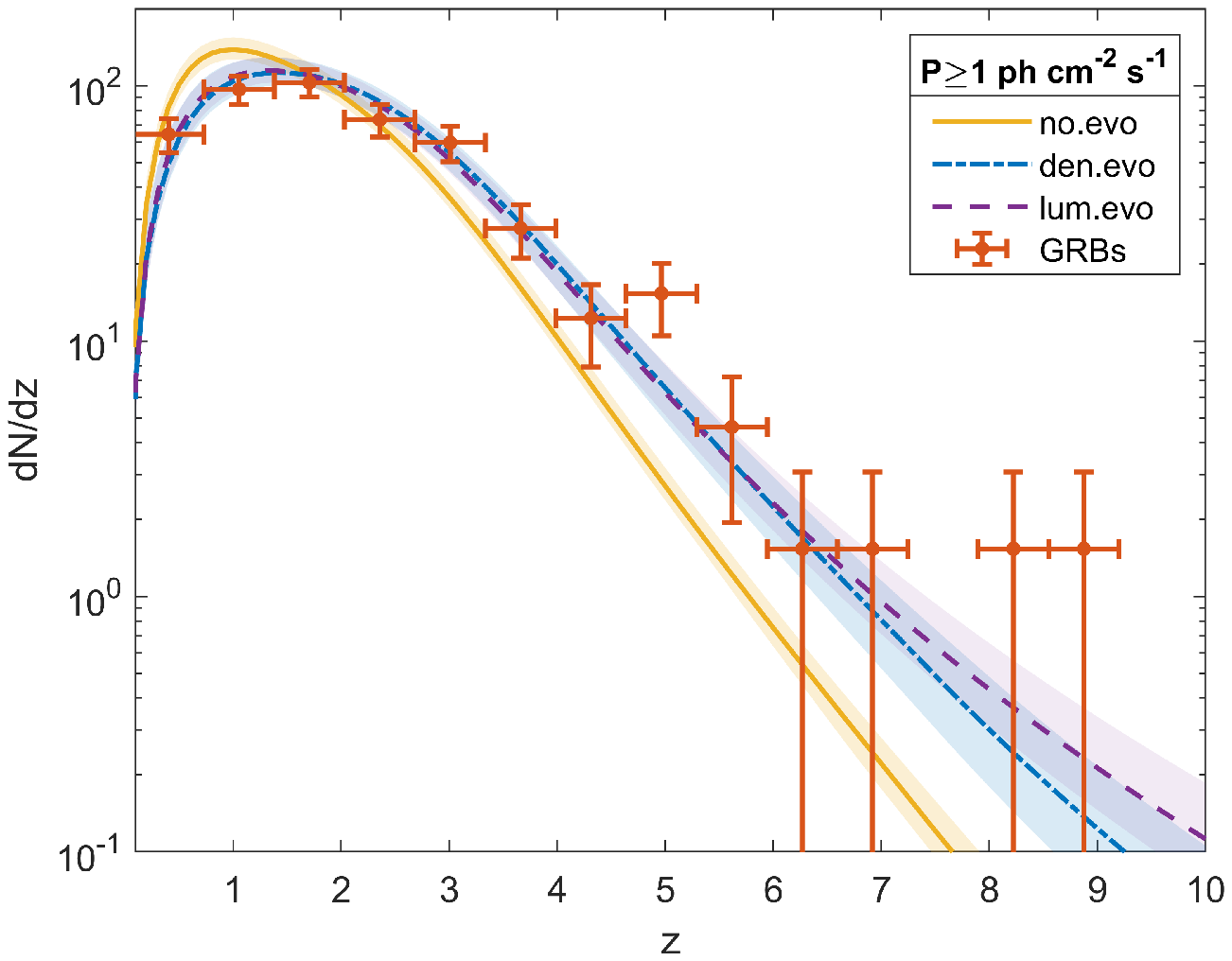} 
\includegraphics[angle=0,scale=0.55]{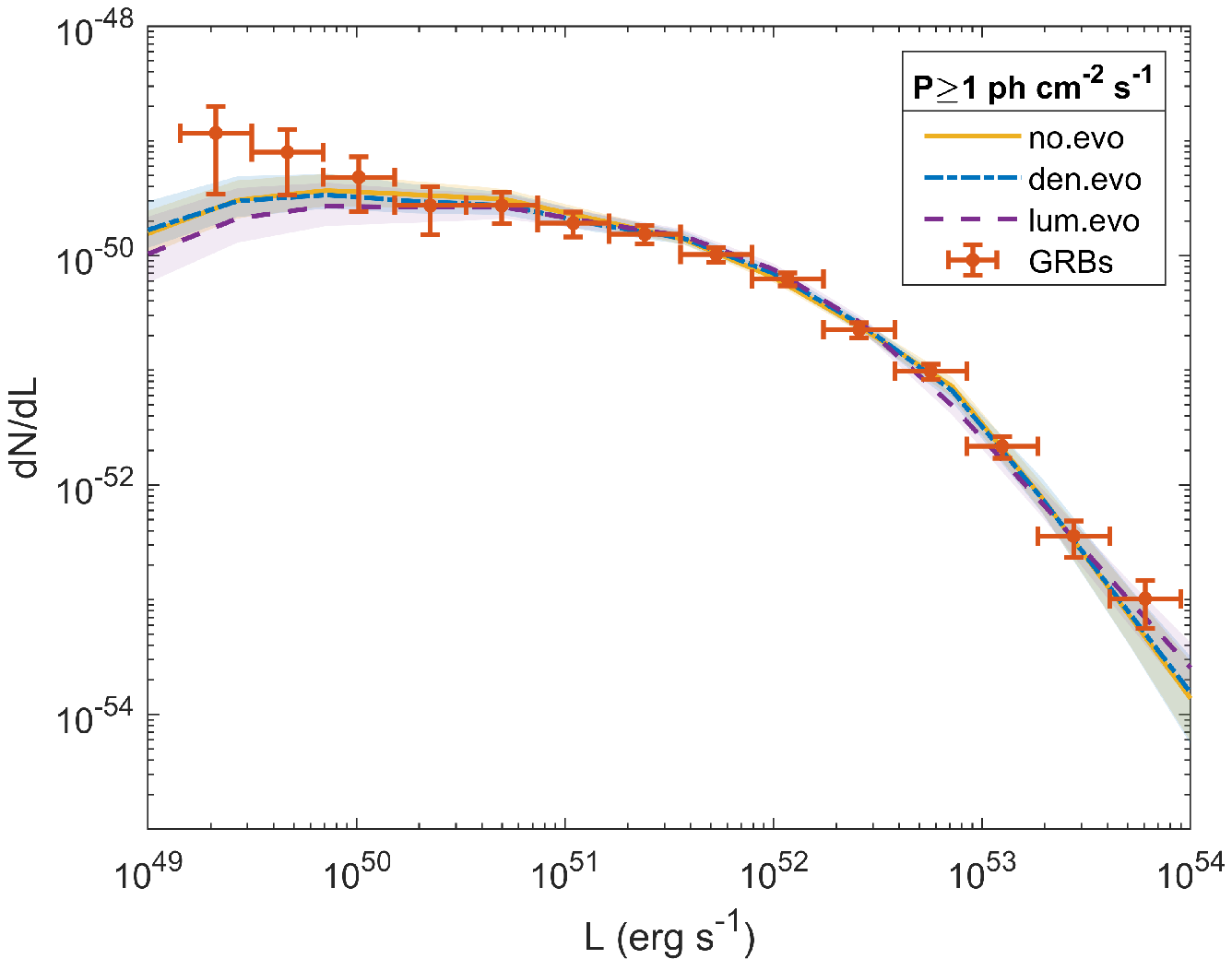}
\caption{Redshift and luminosity distributions of 302 {\it Swift} GRBs with $P\ge1$ photons cm$^{-2}$ s$^{-1}$
(the solid points, with the number of detection in each redshift or luminosity bin indicated by
a red point with Poisson error bars). Different curves correspond to the expected distributions from different
best-fitting models: no evolution model (orange solid lines), density evolution model (blue dot-dashed lines),
and luminosity evolution model (purple dashed lines). Shadow regions represent the $1\sigma$ confidence regions
of the corresponding models. In all models, the broken power-law LF has been assumed.}\label{fig3}
\end{figure*}

\begin{figure*}
\vskip-0.1in
\includegraphics[angle=0,scale=0.55]{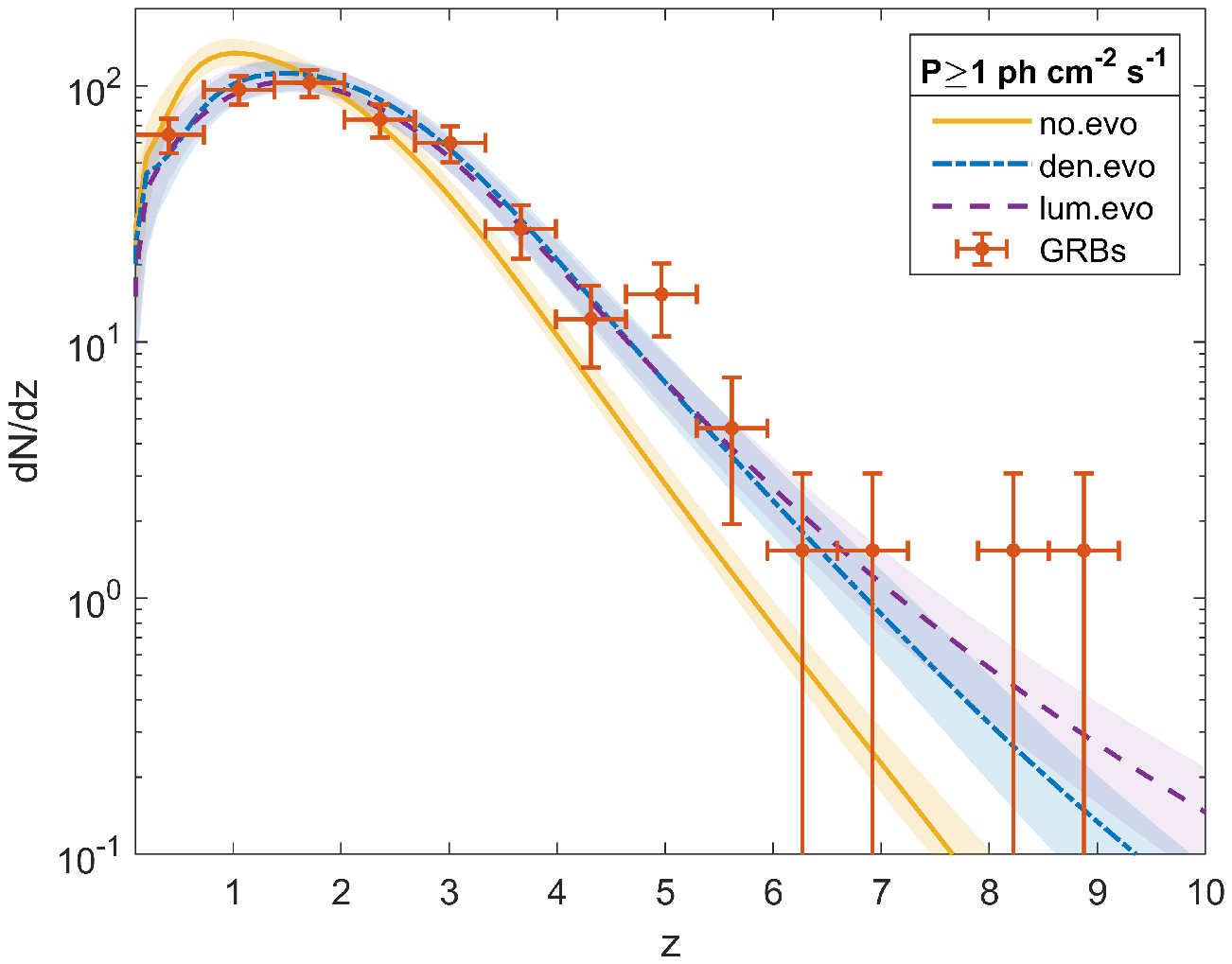} 
\includegraphics[angle=0,scale=0.55]{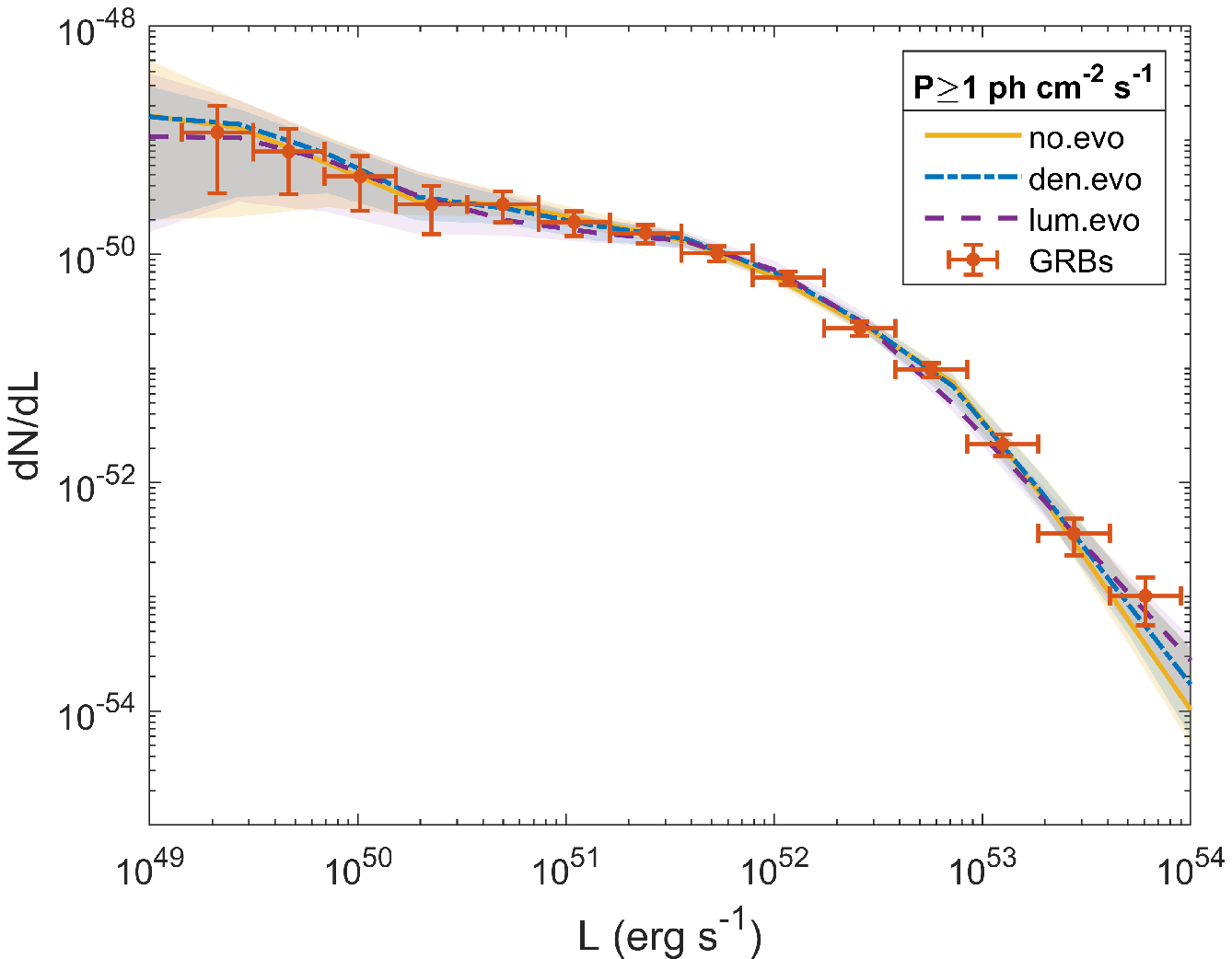}
\caption{Same as Figure~\ref{fig3}, except now for the scenario with the assumed tripe power-law LF.}\label{fig4}
\end{figure*}

The redshift and luminosity distributions of 302 GRBs with $P\geq1$ photons cm$^{-2}$ s$^{-1}$ are shown in
Figure~\ref{fig3} as well as in Figure~\ref{fig4}. We test for each model the two different GRB LF expressions
described in Section~\ref{sec:method}, and we display in Figure~\ref{fig3} the best-fitting results of different
models with the broken power-law LF and in Figure~\ref{fig4} the results with the triple power-law LF.

\subsection{No evolution model}
In the no evolution model, we assume that long GRBs strictly adhere to the star formation history
and that their LF does not evolve with redshift ($L_{c}(z)=L_{c,0}={\rm constant}$).
Figure~\ref{fig3} shows the expected $z$ and $L$ distributions for different models with the assumed broken power-law LF.
It is clear that the expectation from the no evolution model (orange solid lines) does not provide
a good description of the observed distributions of our sample. Particularly, the peak of the expected
$z$ distribution appears at a lower redshift than observed, and both the rate of GRBs at high-\emph{z}
and the $L$ distribution at the low end are underestimated.
On the basis of the AIC model selection criterion, we can safely exclude this model as having
a probability of only $\sim10^{-12}$ of being correct compared to the luminosity evolution model with
the assumed broken power-law LF.

Figure~\ref{fig4} is same as Figure~\ref{fig3}, but with the assumed triple power-law LF.
The results of our fitting from the no evolution model are indicated with orange solid lines. Although the fit of
the $L$ distribution now looks good, the expected $z$ distribution from this model again is incompatible with the
observation. According to the AIC model selection criterion, we confirm that the no evolution model can be
discarded as having a probability of only $\sim10^{-13}$ of being correct compared to the luminosity evolution
model with the assumed triple power-law LF.

\subsection{Luminosity evolution model}
An evolving GRB LF can lead to enhanced detection of GRBs at high-$z$. In this model, the GRB formation rate
still follows the cosmic SFR, but the break luminosity of the GRB LF is assumed to be evolving with redshift
as $L_{c}(z)=L_{c,0}(1+z)^{\delta}$. That is, high-$z$ GRBs are assumed to be brighter than low-$z$ bursts.
With the broken power-law LF, we find that a strong luminosity evolution with $\delta=1.85^{+0.24}_{-0.23}$
reproduces the observed $z$ and $L$ distributions (purple dashed lines in Figure~\ref{fig3}) quite well.
However, we note that the broken power-law LF tends to underestimate the number of low-$L$ bursts with respect
to the observed one.

For the case with the assumed triple power-law LF, both the two break luminosities increase with redshift as
$L_{c_1}(z)=L_{c_{1},0}(1+z)^{\delta}$ and $L_{c_2}(z)=L_{c_{2},0}(1+z)^{\delta}$. We find that a strong
luminosity evolution with $\delta=1.92^{+0.25}_{-0.37}$ is required to reproduce the observed $z$ and $L$
distributions (purple dashed lines in Figure~\ref{fig4}) well. The required evolutionary effect does not
depend on the assumed expression of the GRB LF. However, the AIC shows that the probability of the luminosity
evolution model with the triple power-law LF being correct is $\sim94.3\%$ compared to the same evolution model
but with the broken power-law LF examined above. Note that \cite{2015ApJ...812...33S} also found that the GRB LF
is best characterized as a triple power law.

\subsection{Density evolution model}
An increase of the GRB formation rate with redshift can also enhance the high-$z$ GRB detection. In this model,
the break luminosity in the GRB LF is constant, but the GRB rate is proportional to the cosmic SFR
with an additional evolution parameterized by $(1+z)^{\delta}$, i.e., $\psi(z)=\eta \psi_\star(z)(1+z)^{\delta}$.
With the broken power-law LF, we find that a strong density evolution with $\delta=1.43^{+0.22}_{-0.20}$ fits
the observations best (blue dot-dashed lines in Figure~\ref{fig3}), even though there are still some minor flaws
in the fit of low luminosities.

For the case with the assumed triple power-law LF, the best fit is produced with a density evolution with
$\delta=1.26^{+0.33}_{-0.34} $. This model is represented by the blue dot-dashed lines in Figure~\ref{fig4}.
Obviously, the amount of evolution is independent of the assumed expression of the GRB LF. According to the AIC,
the density evolution model with the triple power-law LF is slightly favoured compared to the same evolution
model with the broken power-law LF, but the differences are statistically insignificant ($\sim61.3\%$
for the former versus $\sim38.7\%$ for the latter).

\section{Extrapolate to whole sample with Trigger probability}
\label{sec:trigger}

\begin{figure}
\vskip-0.1in
  \centering
\includegraphics[angle=0,scale=0.6]{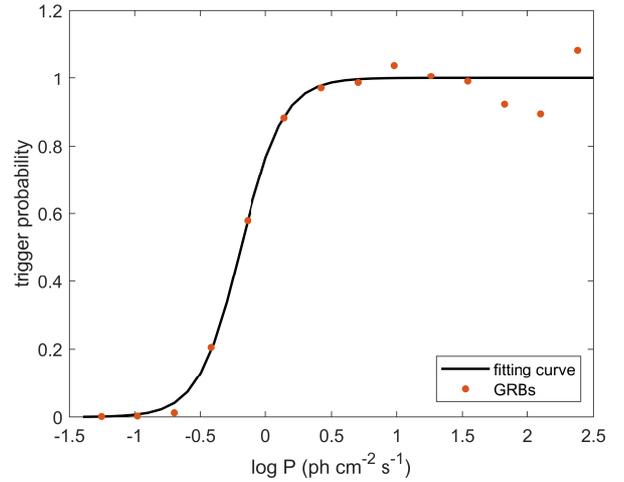}
    \caption{Trigger probability as a function of the peak flux for {\it Swift} long GRBs. The best-fitting curve is parametrized in equation~(\ref{eq:sig}).}\label{fig5}
\end{figure}

The total redshift sample presented in Table~\ref{tabA1} comprises 424 long GRBs. However, due to incomplete sampling of
the faint bursts, only 302 relatively bright bursts with $P\ge1$ photons cm$^{-2}$ s$^{-1}$ have been used in our analysis.
A simple hard cutoff of the trigger threshold, $P_{\rm lim}=1$ photons cm$^{-2}$ s$^{-1}$, has also been assumed, i.e.,
100\% trigger of bursts with $P\geq P_{\rm lim}$. With the inclusion of the other 122 faint bursts with $0.04\leq P < 1$
photons cm$^{-2}$ s$^{-1}$, a more accurate and detailed description of the trigger efficiency $\theta_{\gamma}$ of {\it Swift}/BAT
has to be considered in studying the properties of the long-GRB population. As \cite{2010ApJ...720..435A} did in their
treatment, we estimate the trigger probability of {\it Swift}/BAT through
\begin{equation}
\theta_{\gamma}(P)=\frac{\left({\rm d}N/{\rm d}P\right)_{\rm obs}}{\left({\rm d}N/{\rm d}P\right)_{\rm int}}\;,
\label{eq:TP}
\end{equation}
where $({\rm d}N/{\rm d}P)_{\rm obs}$ is the differential distribution of the observed peak flux (see data points in Figure~\ref{fig1})
and $({\rm d}N/{\rm d}P)_{\rm int}$ is the intrinsic peak-flux distribution. As explained in Section~\ref{subsec:sample}, an approach
to recovering the intrinsic peak-flux distribution is obtained by studying the sample above $P\approx 1$ photons cm$^{-2}$
s$^{-1}$. In order to parameterize the intrinsic peak-flux distribution, we perform a $\chi^{2}$ fit to the differential data
with $P\ge1$ photons cm$^{-2}$ s$^{-1}$ using a broken power-law function:
\begin{equation}
\frac{{\rm d}N}{{\rm d}P} \propto\left\{\begin{array}{l}
P^{\beta_{1}};\;\;\;\;\;\;\;\;\;\;\;\;\;\;\; P \leq P_{c}\;, \\
P_{c}^{\beta_{1}-\beta_{2}}P^{\beta_{2}};\;\;\; P>P_{c}\;,
\end{array}\right.
\label{eq:dNdp}
\end{equation}
where $\beta_{1}$ and $\beta_{2}$ are the power-law indices before and after the break flux $P_{c}$. The best-fitting
broken power-law model with parameters $\beta_{1}=-1.67^{+0.05}_{-0.52}$, $\beta_{2}=-2.37^{+0.72}_{-0.28}$, and
$P_{c}=9.24^{+194.76}_{-0.90}$ photons cm$^{-2}$ s$^{-1}$ is denoted by the green dotted line in Figure~\ref{fig1}.
Figure~\ref{fig1} also shows the expected flux distributions from different best-fitting models with the assumed
tripe power-law LF. Unsurprisingly, the reproduce of the $P$ distribution from the no evolution model (orange solid line)
is not as good as those from the luminosity evolution model (purple dashed line) or density evolution model (blue
dot-dashed line). While, the best-fitting broken power-law model is in good agreement with the expectations from
both the luminosity and density evolution models, implying that the evaluated $({\rm d}N/{\rm d}P)_{\rm int}$ from the broken
power-law model provides a good description of the intrinsic peak-flux distribution. Extrapolating the evaluated
$({\rm d} N/{\rm d}P)_{\rm int}$ to low $P$ ($P<1$ photons cm$^{-2}$ s$^{-1}$), we then estimate the trigger probability
$\theta_{\gamma}(P)$ through Equation~(\ref{eq:TP}). The result is shown in Figure~\ref{fig5}. $\theta_{\gamma}$
as a function of $\log(P)$ can be parametrized by a sigmoid function:
\begin{equation}
\theta_{\gamma}(P)=\frac{1} {1+\exp(\xi_P \log P+\kappa_P)}\;,
\label{eq:sig}
\end{equation}
where the best-fitting parameters are $\xi_P=-6.20\pm0.91$ and $\kappa_P=-1.18\pm0.24$ at 68\% confidence level.

\begin{figure*}
\vskip-0.1in
\includegraphics[angle=0,scale=0.55]{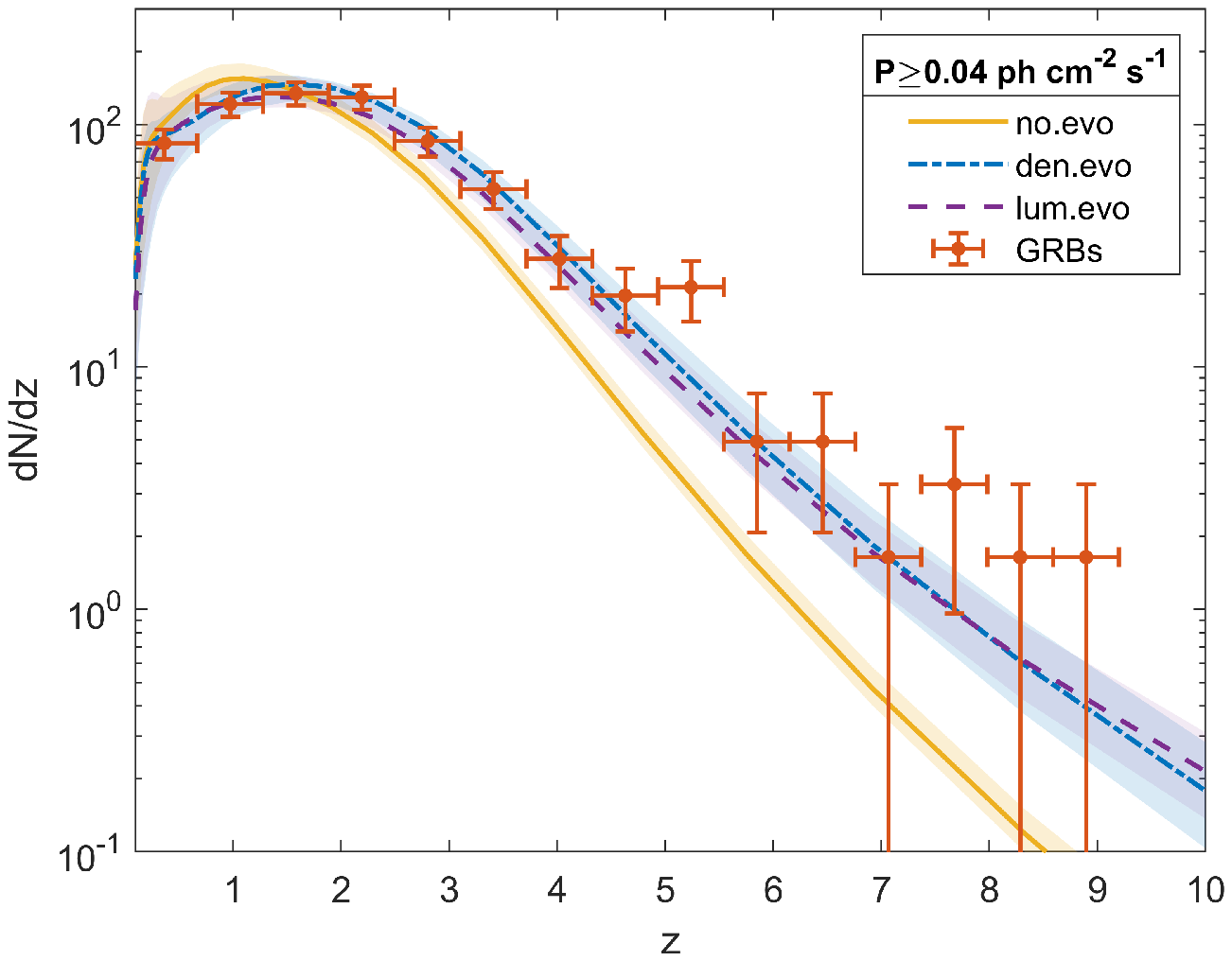} 
\includegraphics[angle=0,scale=0.55]{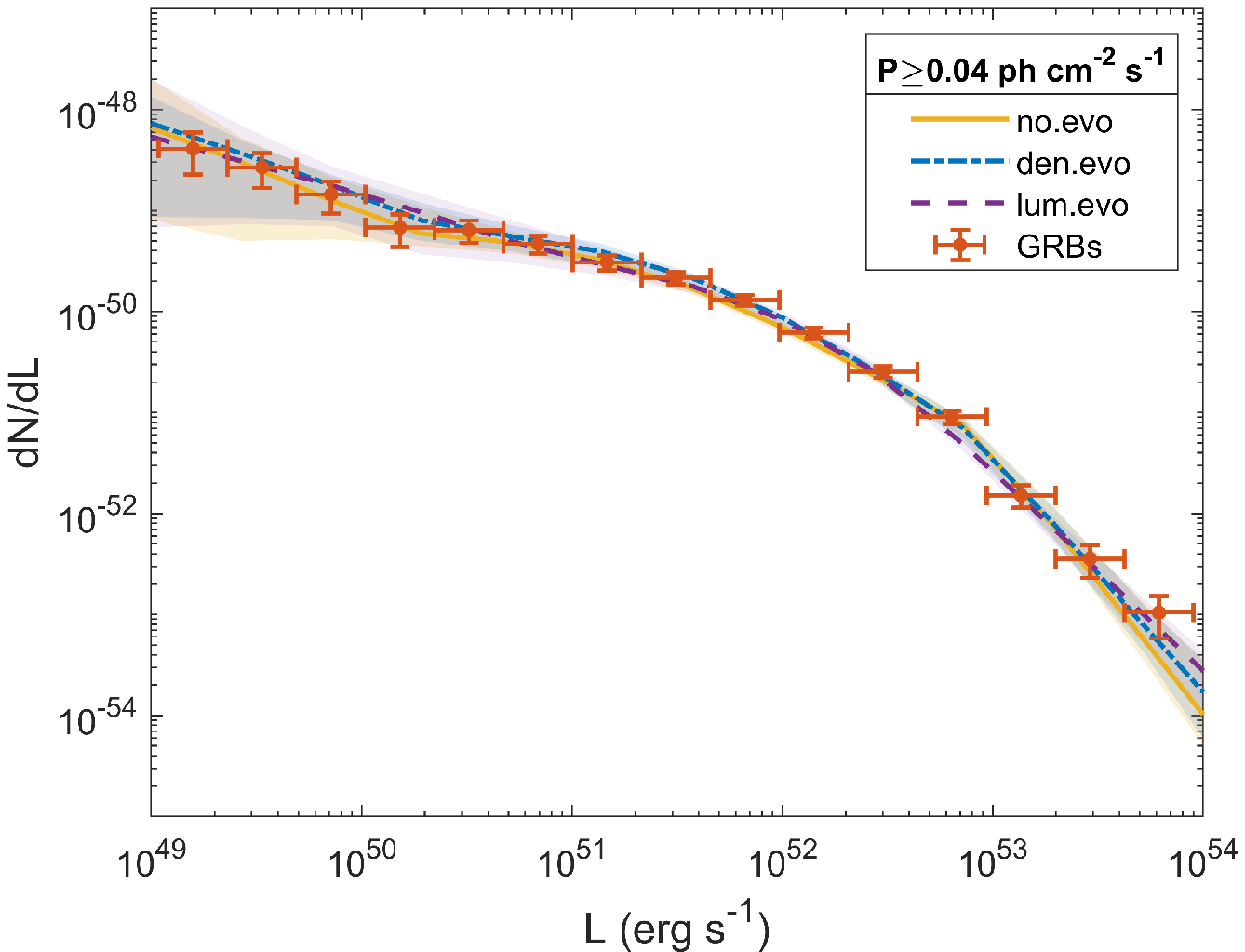}
\caption{Redshift and luminosity distributions of the whole {\it Swift} sample (424 GRBs with $P\ge0.04$ photons cm$^{-2}$ s$^{-1}$).
Model results are shown as in Figure~\ref{fig4}, but now with the inclusion of the trigger probability of {\it Swift}/BAT
(Equation~(\ref{eq:sig})). There has no attempt to fit the observed redshift and luminosity distributions.}\label{fig6}
\end{figure*}

With the evaluated trigger probability of {\it Swift}/BAT (Equation~(\ref{eq:sig})), we then compute the $z$ and
$L$ distributions expected for the whole {\it Swift} redshift sample (424 bursts), by assuming a photon flux limit of
$P_{\rm lim}=0.04$ photons cm$^{-2}$ s$^{-1}$ and fixing the model free parameters to the values given in Table~\ref{tab1}
(entries 4--6, corresponding to the case with the assumed triple power-law LF). The results are shown in Figure~\ref{fig6}
for the different evolution models explored here. The models are compared with the $z$ and $L$ distributions
inferred from the whole redshift sample of GRBs detected by {\it Swift}. This whole sample has a redshift completeness level
similar to our flux-cut sample but is larger in size and covers a broader flux range. We find that the expectations
from our evolution models offer a good representation of the observed $z$ and $L$ distributions of the whole sample
without the need for any adjustment of the model free parameters, whereas the no evolution model gives a poor match.
This further supports the reliability of our analysis and sharpens the strength of our conclusions.


\section{Conclusions}
\label{sec:summary}

In this work, we update and enlarge the long GRB sample detected by the {\it Swift} satellite. The enlarged sample contains
1111 GRBs up to 2019 November, among which 424 bursts have measured redshift. Given the incomplete sampling of the faint
bursts, we restrict ourselves to GRBs that are relatively bright in the 15--150 keV {\it Swift}/BAT energy band. In practice,
we select bursts having 1-s peak photon flux $P\geq1$ photons cm$^{-2}$ s$^{-1}$. Above this flux limit, the
incomplete flux sampling induced by the instrumental selection effect becomes negligible. Our subsample is then composed of
733 GRBs with $P\geq1$ photons cm$^{-2}$ s$^{-1}$, 302 of them with known redshifts for a completeness level of
41\%. Given the low completeness level in redshift measurement, we model the redshift measurement probability based on
the current {\it Swift} GRB sample.

Using the flux-limited GRB subsample, we construct the GRB LF and redshift distribution in the frameworks of different
evolution models by taking into account the redshift measurement probability. The maximum likelihood algorithm is applied
to perform an analysis for the redshift and luminosity distributions of the GRB sample, and the MCMC technology is applied
to obtain the best-fitting parameters in each model. We find that GRBs must have experienced some sort of evolution with
redshift, being more luminous or more numerous in the past than present day. In order to reproduce the observed distributions
well, the burst luminosity should increase to $(1+z)^{1.92^{+0.25}_{-0.37}}$ or the GRB rate density should rise to
$(1+z)^{1.26^{+0.33}_{-0.34}}$ with respect to the known cosmic evolution of the SFR, confirming previous findings
derived from the similar parametric studies (e.g., \citealt{2012ApJ...749...68S,2019MNRAS.488.4607L}).
Besides these parametric studies, the non-parametric statistical method proposed by \cite{1971MNRAS.155...95L} and \cite{1992ApJ...399..345E} has also been widely used to determine the evolution of the GRB LF (e.g.,
\citealt{2002ApJ...574..554L,2015ApJ...806...44P,2015ApJS..218...13Y,2016A&A...587A..40P}). To our knowledge,
the first attempt to determine the GRB luminosity evolution through the non-parametric method was presented
in \cite{2002ApJ...574..554L}. We would like to stress that if we focus on the luminosity evolution model,
our derived evolution index ($\delta=1.92^{+0.25}_{-0.37}$) is broadly consistent with previous results based on the
non-parametric studies: $\delta=1.4\pm0.5$ by \cite{2002ApJ...574..554L}, $\delta=2.3\pm0.8$ by \cite{2015ApJ...806...44P},
$\delta=2.43^{+0.41}_{-0.38}$ by \cite{2015ApJS..218...13Y}, and $\delta\sim2.5$ by \cite{2016A&A...587A..40P}.

We also explore two general expressions for the GRB LF: a broken power-law
LF and a triple power-law LF. We find that the required evolutionary effect does not depend on the assumed GRB LF expression.
However, in a one-on-one comparison using the AIC model selection criterion, the luminosity evolution model involving
the triple power-law LF is statistically preferred over the same model but involving the broken power-law LF with a relative
probability of $\sim94.3$\% versus $\sim5.7$\%. \cite{2015ApJ...812...33S} also suggested that a triple power-law expression
can give the best description for the GRB LF.

Extrapolating our fitting results to the flux limit ($P=0.04$ photons cm$^{-2}$ s$^{-1}$) of the whole {\it Swift} redshift sample,
and considering the trigger probability of {\it Swift}/BAT in detail, we find that the expectations from our luminosity and density
evolution models are well consistent with the observations of the whole sample without the requirement for any adjustment of the model free parameters. This further confirms that our analysis results are reliable. We also note that
since the two evolution models explored here predict very similar distributions, it is difficult to distinguish between luminosity and density evolution only on the basis of the current {\it Swift} sample.

\section*{Acknowledgements}
We are grateful to the anonymous referee for insightful comments.
This work is partially supported by the National Natural Science Foundation of China
(grant Nos. 11725314, U1831122, 12041306 and 11703094), the Youth Innovation Promotion
Association (2017366), the Key Research Program of Frontier Sciences (grant No.
ZDBS-LY-7014) of Chinese Academy of Sciences, and the Major Science and Technology Project
of Qinghai Province (2019-ZJ-A10).

\section*{Data availability}
The data underlying this article will be shared on reasonable request to the corresponding author.







\appendix

\section{Additional table}


\onecolumn
\begin{longtable}{lccccccc}
\caption{List of the bursts with known redshifts}
\label{tabA1} \\ 
\hline
GRB & \tabincell{c}{Peak flux \\(photons cm$^{-2}$ s$^{-1})$} & $z$  &$\alpha$ & $\beta$ &  \tabincell{c}{$E_p$\\(keV)} & \tabincell{c}{$\log L$ \\(erg s$^{-1}$)}& Refs. \\
\hline
\endfirsthead

\hline
GRB & \tabincell{c}{Peak flux \\(photons cm$^{-2}$ s$^{-1})$} & $z$  &$\alpha$ & $\beta$ &  \tabincell{c}{$E_p$\\(keV)} & \tabincell{c}{$\log L$ \\(erg s$^{-1}$)}& Refs. \\
\hline
\endhead

\hline
\endfoot
041228&$1.61\pm0.25$&$2.3$&$-1.54 ^{+0.05 }_{-0.05 }$&$$&$240.00 ^{+381.06 }_{-66.53 }$&$52.09 \pm0.07 $&$3,3$\\
050126&$0.71\pm0.17$&$1.29$&$-1.10 ^{+0.10 }_{-0.10 }$&$-2.20 $&$47.00 ^{+23.00 }_{-8.00 }$&$51.02 \pm0.10 $&$1,3$\\
050215B&$0.67\pm0.12$&$2.62$&$-1.10 $&$-2.20 ^{+0.60 }_{-0.40 }$&$17.60 ^{+6.10 }_{-12.80 }$&$51.74 \pm0.08 $&$1,3$\\
050219A&$3.53\pm0.35$&$0.211£¿$&$0.00 ^{+0.18 }_{-0.18 }$&$$&$105.00 ^{+10.28 }_{-6.65 }$&$49.74 \pm0.04 $&$1,3$\\
050223&$0.69\pm0.16$&$0.5915$&$-1.74 ^{+0.17 }_{-0.18 }$&$$&$67.03 ^{+113.11 }_{-22.99 }$&$50.15 \pm0.10 $&$1,4$\\
050315&$1.93\pm0.22$&$1.949$&$-1.10 $&$-2.04 ^{+0.16 }_{-0.24 }$&$40.30 ^{+8.50 }_{-11.40 }$&$51.96 \pm0.05 $&$1,3$\\
050318&$3.16\pm0.20$&$1.44$&$-1.22 ^{+0.45 }_{-0.41 }$&$$&$50.86 ^{+12.24 }_{-7.17 }$&$51.60 \pm0.03 $&$1,4$\\
050319&$1.52\pm0.21$&$3.24$&$-2.00 ^{+0.19 }_{-0.21 }$&$$&$44.73 ^{+27.49 }_{-43.11 }$&$52.54 \pm0.06 $&$1,4$\\
050401&$10.70\pm0.92$&$2.9$&$-1.44 ^{+0.08 }_{-0.08 }$&$$&$165.49 ^{+391.26 }_{-48.08 }$&$53.09 \pm0.04 $&$1,4$\\
050406&$0.36\pm0.10$&$2.44$&$-1.10 ^{+0.40 }_{-0.40 }$&$-2.56 ^{+0.35 }_{-0.35 }$&$25.00 ^{+35.00 }_{-13.00 }$&$51.28 \pm0.12 $&$2,3$\\
050410&$1.80\pm0.36$&$1.04$&$-0.90 ^{+0.24 }_{-0.24 }$&$$&$92.00 ^{+32.06 }_{-13.31 }$&$51.06 \pm0.09 $&$3,3$\\
050412&$0.48\pm0.05$&$4.5$&$-0.60 ^{+0.12 }_{-0.12 }$&$$&$640.00 ^{+574.62 }_{-199.60 }$&$52.86 \pm0.05 $&$3,3$\\
050416A&$4.88\pm0.48$&$0.6535$&$-0.97 ^{+2.28 }_{-1.01 }$&$$&$13.67 ^{+7.93 }_{-12.56 }$&$51.00 \pm0.04 $&$2,3$\\
050502B&$1.42\pm0.13$&$5.2$&$-1.60 ^{+0.06 }_{-0.06 }$&$$&$100.00 ^{+228.03 }_{-19.96 }$&$52.78 \pm0.04 $&$1,3$\\
050505&$1.85\pm0.31$&$4.27$&$-1.41 ^{+0.12 }_{-0.12 }$&$$&$140.25 ^{+343.09 }_{-42.77 }$&$52.70 \pm0.07 $&$1,4$\\
050525A&$41.70\pm0.94$&$0.606$&$-1.10 ^{+0.03 }_{-0.03 }$&$$&$84.10 ^{+1.03 }_{-1.03 }$&$51.84 \pm0.01 $&$2,3$\\
050603&$21.50\pm1.07$&$2.821$&$-1.03 ^{+0.11 }_{-0.11 }$&$-2.03 ^{+0.17 }_{-0.29 }$&$343.70 ^{+87.00 }_{-87.00 }$&$53.73 \pm0.02 $&$3,3$\\
050607&$0.95\pm0.14$&$4$&$-1.80 ^{+0.12 }_{-0.12 }$&$$&$67.00 ^{+138.51 }_{-16.94 }$&$52.39 \pm0.06 $&$3,3$\\
050714B&$0.52\pm0.22$&$2.4383$&$-0.42 ^{+1.93 }_{-1.43 }$&$$&$28.71 ^{+7.26 }_{-28.70 }$&$51.24 \pm0.18 $&$1,3$\\
050716&$2.18\pm0.35$&$1.4$&$-0.80 ^{+0.18 }_{-0.18 }$&$$&$136.00 ^{+61.70 }_{-18.15 }$&$51.57 \pm0.07 $&$3,3$\\
050724&$3.26\pm0.30$&$0.257$&$-1.66 ^{+0.17 }_{-0.18 }$&$$&$100.00 ^{+376.82 }_{-40.98 }$&$49.98 \pm0.04 $&$2,4$\\
050730&$0.55\pm0.14$&$3.967$&$-1.39 ^{+0.13 }_{-0.13 }$&$$&$164.87 ^{+499.25 }_{-59.36 }$&$52.12 \pm0.11 $&$1,4$\\
050801&$1.46\pm0.14$&$1.38$&$-1.90 ^{+0.12 }_{-0.12 }$&$$&$44.00 ^{+35.08 }_{-25.40 }$&$51.46 \pm0.04 $&$1,3$\\
050802&$2.75\pm0.44$&$1.71£¿$&$-1.59 ^{+0.14 }_{-0.15 }$&$$&$95.12 ^{+177.17 }_{-31.15 }$&$51.88 \pm0.07 $&$1,4$\\
050803&$0.96\pm0.11$&$3.5$&$-1.30 ^{+0.06 }_{-0.06 }$&$$&$235.00 ^{+382.27 }_{-56.86 }$&$52.31 \pm0.05 $&$1,3$\\
050814&$0.71\pm0.25$&$5.77$&$-0.58 ^{+0.56 }_{-0.56 }$&$-2.20 $&$54.00 ^{+7.50 }_{-7.50 }$&$52.55 \pm0.15 $&$1,3$\\
050819&$0.38\pm0.12$&$2.5043$&$-0.65 ^{+2.62 }_{-1.34 }$&$$&$20.19 ^{+10.66 }_{-17.49 }$&$51.20 \pm0.14 $&$1,4$\\
050820A&$2.45\pm0.23$&$2.612$&$-1.20 ^{+0.06 }_{-0.06 }$&$$&$490.00 ^{+435.50 }_{-181.46 }$&$52.64 \pm0.04 $&$2,3$\\
050822&$2.24\pm0.22$&$1.434$&$-1.00 ^{*}$&$-2.21 ^{+0.14 }_{-0.14 }$&$27.25 ^{+7.09 }_{-25.74 }$&$51.60 \pm0.04 $&$1,4$\\
050824&$0.50\pm0.15$&$0.83$&$-1.00 ^{*}$&$-2.91 ^{+0.43 }_{-0.53 }$&$13.85 ^{+2.74 }_{-12.85 }$&$50.29 \pm0.13 $&$1,4$\\
050826&$0.38\pm0.13$&$0.297$&$-1.22 ^{+0.27 }_{-0.27 }$&$$&$340.31 ^{+792.76 }_{-209.24 }$&$49.42 \pm0.15 $&$1,4$\\
050904&$0.62\pm0.17$&$6.29$&$-1.17 ^{+0.07 }_{-0.07 }$&$$&$531.78 ^{+775.65 }_{-297.30 }$&$52.95 \pm0.12 $&$1,4$\\
050908&$0.70\pm0.14$&$3.344$&$-1.10 $&$-2.20 $&$41.00 ^{+9.00 }_{-5.00 }$&$51.99 \pm0.09 $&$1,3$\\
050911&$1.33\pm0.20$&$0.165$&$-1.80 ^{+0.24 }_{-0.24 }$&$$&$55.00 ^{+218.36 }_{-31.45 }$&$49.15 \pm0.07 $&$3,3$\\
050915A&$0.77\pm0.14$&$2.5273$&$-1.30 ^{+0.12 }_{-0.12 }$&$$&$179.00 ^{+321.18 }_{-38.11 }$&$51.81 \pm0.08 $&$2,3$\\
050922B&$1.01\pm0.36$&$4.5$&$-1.10 $&$-2.10 ^{+0.18 }_{-0.18 }$&$38.00 ^{+15.73 }_{-22.38 }$&$52.49 \pm0.15 $&$1,3$\\
050922C&$7.26\pm0.32$&$2.198$&$-0.83 ^{+0.23 }_{-0.26 }$&$$&$130.00 ^{+51.00 }_{-27.00 }$&$52.56 \pm0.02 $&$1,3$\\
051001&$0.49\pm0.11$&$2.4296$&$-1.12 ^{+0.65 }_{-0.56 }$&$$&$44.38 ^{+11.49 }_{-8.01 }$&$51.32 \pm0.10 $&$1,4$\\
051006&$1.62\pm0.30$&$1.059$&$-1.40 ^{+0.12 }_{-0.12 }$&$$&$270.00 ^{+350.82 }_{-102.83 }$&$51.29 \pm0.08 $&$1,3$\\
051008&$5.44\pm0.35$&$2.77$&$-0.80 ^{+0.23 }_{-0.22 }$&$$&$234.76 ^{+315.18 }_{-70.12 }$&$52.89 \pm0.03 $&$1,4$\\
051016B&$1.30\pm0.16$&$0.9364$&$-1.10 $&$-2.56 ^{+0.50 }_{-6.33 }$&$27.50 ^{+13.20 }_{-12.30 }$&$50.81 \pm0.05 $&$1,3$\\
051021B&$0.80\pm0.14$&$2.1$&$-0.40 ^{+0.48 }_{-0.48 }$&$$&$99.00 ^{+73.79 }_{-19.36 }$&$51.47 \pm0.08 $&$3,3$\\
051109A&$3.94\pm0.69$&$2.346$&$-1.25 ^{+0.44 }_{-0.59 }$&$  $&$161.00 ^{+224.00 }_{-58.00 }$&$52.42 \pm0.08 $&$2,3$\\
051111&$2.66\pm0.21$&$1.55$&$-1.22 ^{+0.09 }_{-0.09 }$&$-2.10 ^{+4.94 }_{-0.27 }$&$447.00 ^{+206.00 }_{-175.00 }$&$52.19 \pm0.03 $&$1,3$\\
051117B&$0.49\pm0.14$&$0.481$&$-1.70 ^{+0.24 }_{-0.24 }$&$$&$72.00 ^{+257.67 }_{-22.38 }$&$49.78 \pm0.12 $&$1,3$\\
051227&$0.95\pm0.12$&$0.8$&$-1.10 ^{+0.12 }_{-0.12 }$&$$&$340.00 ^{+387.11 }_{-127.02 }$&$50.88 \pm0.05 $&$3,3$\\
060108&$0.77\pm0.12$&$2$&$-1.10 $&$-2.01 ^{+0.01 }_{-0.12 }$&$44.00 ^{+16.94 }_{-25.40 }$&$51.61 \pm0.07 $&$1,3$\\
060110&$1.89\pm0.16$&$5$&$-1.58 ^{+0.05 }_{-0.05 }$&$$&$135.00 ^{+210.49 }_{-28.43 }$&$52.89 \pm0.04 $&$3,3$\\
060111A&$1.71\pm0.15$&$2.32$&$-0.80 ^{+0.24 }_{-0.24 }$&$-2.29 ^{+0.15 }_{-0.19 }$&$83.00 ^{+14.52 }_{-7.26 }$&$52.06 \pm0.04 $&$1,3$\\
060111B&$1.40\pm0.27$&$1.5$&$-0.90 ^{+0.12 }_{-0.12 }$&$-2.35 ^{+0.36 }_{-4.63 }$&$540.00 ^{+568.57 }_{-169.36 }$&$52.05 \pm0.08 $&$3,3$\\
060115&$0.87\pm0.12$&$3.53$&$-1.15 ^{+0.58 }_{-0.51 }$&$$&$69.81 ^{+121.75 }_{-15.07 }$&$52.01 \pm0.06 $&$1,4$\\
060123&$0.04$&$0.56$&$-1.90 ^{+0.60 }_{-0.60 }$&$$&$250.00 ^{*}$&$49.03 \pm0.00 $&$1,3$\\
060124&$0.89\pm0.18$&$2.296$&$5.00 ^{+0.00 }_{-2.89 }$&$-2.00 ^{+0.00 }_{-0.15 }$&$29.64 ^{+3.67 }_{-4.31 }$&$51.79 \pm0.09 $&$1,4$\\
060202&$0.51\pm0.17$&$0.783$&$-1.74 ^{+0.13 }_{-0.13 }$&$$&$87.72 ^{+344.22 }_{-26.36 }$&$50.34 \pm0.14 $&$1,4$\\
060204B&$1.35\pm0.15$&$2.3393$&$-0.70 ^{+0.18 }_{-0.18 }$&$$&$125.00 ^{+47.18 }_{-17.54 }$&$51.88 \pm0.05 $&$1,3$\\
060206&$2.79\pm0.17$&$4.048$&$-1.18 ^{+0.33 }_{-0.31 }$&$$&$87.03 ^{+53.26 }_{-15.82 }$&$52.70 \pm0.03 $&$1,4$\\
060210&$2.72\pm0.28$&$3.91$&$-1.47 ^{+0.10 }_{-0.10 }$&$$&$136.23 ^{+347.11 }_{-38.75 }$&$52.78 \pm0.04 $&$1,4$\\
060223A&$1.35\pm0.18$&$4.41$&$-1.18 ^{+0.31 }_{-0.31 }$&$-2.20 $&$71.00 ^{+100.00 }_{-10.00 }$&$52.60 \pm0.06 $&$2,3$\\
060306&$5.97\pm0.35$&$1.559$&$-1.13 ^{+0.47 }_{-0.43 }$&$$&$67.17 ^{+43.83 }_{-11.66 }$&$51.98 \pm0.03 $&$1,4$\\
060319&$1.09\pm0.14$&$1.172$&$-1.00 ^{*}$&$-2.19 ^{+0.23 }_{-0.26 }$&$24.66 ^{+3.59 }_{-23.66 }$&$51.09 \pm0.06 $&$1,4$\\
060403&$0.95\pm0.13$&$1.3$&$-0.40 ^{+0.24 }_{-0.24 }$&$$&$247.00 ^{+239.53 }_{-58.07 }$&$51.43 \pm0.06 $&$3,3$\\
060418&$6.52\pm0.35$&$1.489$&$-1.61 ^{+0.05 }_{-0.05 }$&$$&$134.99 ^{+276.23 }_{-33.35 }$&$52.16 \pm0.02 $&$1,4$\\
060428B&$0.66\pm0.15$&$0.348$&$-0.80 ^{+0.97 }_{-0.97 }$&$-2.20 ^{+0.40 }_{-0.40 }$&$23.00 ^{+3.00 }_{-8.00 }$&$49.59 \pm0.10 $&$3,3$\\
060501&$1.98\pm0.26$&$1.8$&$-1.20 ^{+0.06 }_{-0.06 }$&$$&$217.00 ^{+285.50 }_{-52.62 }$&$51.91 \pm0.06 $&$3,3$\\
060502A&$1.69\pm0.21$&$~1.51$&$-0.87 ^{+0.21 }_{-0.15 }$&$-1.98 ^{+0.16 }_{-1.01 }$&$174.00 ^{+87.10 }_{-36.29 }$&$51.90 \pm0.05 $&$3,3$\\
060505&$2.65\pm0.63$&$0.089$&$$&$$&$$&$49.23 \pm0.10 $&$1,$\\
060510A&$14.70\pm1.13$&$1.2$&$-1.80 ^{+0.06 }_{-0.06 }$&$$&$74.00 ^{+146.98 }_{-17.54 }$&$52.28 \pm0.03 $&$3,3$\\
060510B&$0.57\pm0.11$&$4.94$&$-1.47 ^{+0.18 }_{-0.18 }$&$-2.20 $&$95.00 ^{+60.00 }_{-30.00 }$&$52.38 \pm0.08 $&$1,3$\\
060512&$0.88\pm0.20$&$2.1$&$-0.50 ^{+3.80 }_{-1.49 }$&$$&$22.31 ^{+11.50 }_{-19.51 }$&$51.34 \pm0.10 $&$1,4$\\
060522&$0.55\pm0.15$&$5.11$&$-0.70 ^{+0.44 }_{-0.44 }$&$-2.20 $&$70.00 ^{+13.00 }_{-13.00 }$&$52.36 \pm0.12 $&$1,3$\\
060526&$1.67\pm0.18$&$3.221$&$-1.88 ^{+0.24 }_{-0.25 }$&$$&$73.25 ^{+151.30 }_{-72.25 }$&$52.45 \pm0.05 $&$1,4$\\
060602A&$0.56\pm0.20$&$0.787$&$-1.30 ^{+0.12 }_{-0.12 }$&$$&$280.00 ^{+344.77 }_{-90.73 }$&$50.53 \pm0.16 $&$2,3$\\
060604&$0.34\pm0.13$&$2.1357$&$-1.00 ^{*}$&$-2.17 ^{+0.54 }_{-0.69 }$&$36.82 ^{+49.21 }_{-35.82 }$&$51.22 \pm0.17 $&$1,4$\\
060605&$0.46\pm0.12$&$3.78$&$-0.29 ^{+0.88 }_{-0.73 }$&$$&$87.04 ^{+91.78 }_{-19.90 }$&$51.80 \pm0.11 $&$1,4$\\
060607A&$1.40\pm0.13$&$3.082$&$-0.60 ^{+0.70 }_{-0.60 }$&$$&$63.00 ^{+21.00 }_{-10.00 }$&$52.00 \pm0.04 $&$2,3$\\
060614&$11.50\pm0.74$&$0.125$&$-1.90 ^{+0.04 }_{-0.04 }$&$$&$393.02 ^{+818.57 }_{-250.96 }$&$50.03 \pm0.03 $&$1,4$\\
060707&$1.01\pm0.23$&$3.425$&$-0.66 ^{+0.79 }_{-0.67 }$&$$&$60.94 ^{+22.69 }_{-9.27 }$&$51.97 \pm0.10 $&$1,4$\\
060708&$1.94\pm0.14$&$1.92$&$-1.00 ^{+0.30 }_{-0.30 }$&$$&$99.00 ^{+73.79 }_{-17.54 }$&$51.77 \pm0.03 $&$1,3$\\
060714&$1.28\pm0.13$&$2.711$&$-1.90 ^{+0.11 }_{-0.11 }$&$$&$53.17 ^{+94.49 }_{-49.44 }$&$52.16 \pm0.04 $&$1,4$\\
060719&$2.16\pm0.20$&$1.532$&$-1.93 ^{+0.11 }_{-0.11 }$&$$&$54.88 ^{+44.95 }_{-53.22 }$&$51.78 \pm0.04 $&$1,4$\\
060729&$1.17\pm0.13$&$0.54$&$-1.77 ^{+0.14 }_{-0.14 }$&$$&$67.03 ^{+140.90 }_{-28.43 }$&$50.30 \pm0.05 $&$1,4$\\
060805A&$0.34\pm0.11$&$2.3633$&$5.00 ^{+0.00 }_{-4.46 }$&$$&$30.27 ^{+5.19 }_{-4.39 }$&$50.85 \pm0.14 $&$1,4$\\
060814&$7.27\pm0.29$&$1.9229$&$-1.29 ^{+0.16 }_{-0.15 }$&$$&$238.90 ^{+891.04 }_{-83.43 }$&$52.56 \pm0.02 $&$1,4$\\
060904B&$2.44\pm0.21$&$0.703$&$-0.61 ^{+0.42 }_{-0.42 }$&$-1.78 ^{+0.23 }_{-0.16 }$&$103.00 ^{+59.00 }_{-26.00 }$&$51.34 \pm0.04 $&$1,3$\\
060906&$1.97\pm0.28$&$3.686$&$-1.97 ^{+0.14 }_{-0.14 }$&$$&$46.56 ^{+24.61 }_{-44.93 }$&$52.73 \pm0.06 $&$1,4$\\
060908&$3.03\pm0.25$&$1.8836$&$-0.73 ^{+0.28 }_{-0.26 }$&$$&$127.72 ^{+61.65 }_{-24.75 }$&$52.01 \pm0.04 $&$1,4$\\
060912A&$8.58\pm0.44$&$0.937£¿$&$-1.70 ^{+0.05 }_{-0.05 }$&$$&$200.00 ^{+338.72 }_{-66.53 }$&$51.84 \pm0.02 $&$2,3$\\
060923B&$1.52\pm0.33$&$1.5094$&$-1.00 ^{*}$&$-2.61 ^{+0.29 }_{-0.33 }$&$21.38 ^{+9.24 }_{-19.33 }$&$51.38 \pm0.09 $&$1,4$\\
060926&$1.09\pm0.14$&$3.2$&$-1.00 ^{*}$&$-2.49 ^{+0.23 }_{-0.25 }$&$20.19 ^{+7.70 }_{-18.73 }$&$52.06 \pm0.06 $&$1,4$\\
060927&$2.70\pm0.17$&$5.47$&$-0.82 ^{+0.43 }_{-0.39 }$&$$&$70.90 ^{+21.82 }_{-9.94 }$&$52.92 \pm0.03 $&$1,4$\\
061004&$2.54\pm0.16$&$3.3$&$-1.79 ^{+0.06 }_{-0.06 }$&$$&$64.00 ^{+25.40 }_{-13.31 }$&$52.60 \pm0.03 $&$3,3$\\
061006A&$5.24\pm0.21$&$0.438$&$-0.62 ^{+0.18 }_{-0.21 }$&$$&$664.00 ^{+227.00 }_{-144.00 }$&$51.47 \pm0.02 $&$3,3$\\
061007&$14.60\pm0.37$&$1.261$&$-0.96 ^{+0.03 }_{-0.03 }$&$$&$838.02 ^{+1099.77 }_{-326.78 }$&$52.95 \pm0.01 $&$1,4$\\
061021&$6.11\pm0.27$&$0.3463$&$-1.27 ^{+0.05 }_{-0.05 }$&$$&$405.52 ^{+806.07 }_{-167.84 }$&$50.82 \pm0.02 $&$1,4$\\
061028&$0.65\pm0.19$&$0.76$&$-1.70 ^{+0.18 }_{-0.18 }$&$$&$79.00 ^{+262.51 }_{-23.59 }$&$50.39 \pm0.13 $&$3,3$\\
061110A&$0.53\pm0.12$&$0.758$&$-1.60 ^{+0.06 }_{-0.06 }$&$$&$106.00 ^{+181.46 }_{-24.19 }$&$50.30 \pm0.10 $&$2,3$\\
061110B&$0.45\pm0.11$&$3.44$&$-0.70 ^{+0.12 }_{-0.12 }$&$-4.24 ^{+1.42 }_{-3.48 }$&$550.00 ^{+592.77 }_{-151.22 }$&$52.45 \pm0.11 $&$1,3$\\
061121&$21.10\pm0.46$&$1.314$&$-1.31 ^{+0.02 }_{-0.02 }$&$-2.37 ^{+0.19 }_{-0.73 }$&$741.00 ^{+129.44 }_{-102.22 }$&$52.95 \pm0.01 $&$1,3$\\
061126&$9.76\pm0.38$&$1.1588$&$-1.16 ^{+0.02 }_{-0.02 }$&$$&$390.00 ^{+381.06 }_{-151.22 }$&$52.31 \pm0.02 $&$1,3$\\
061202&$2.51\pm0.17$&$2.2543$&$-1.20 ^{+0.33 }_{-0.31 }$&$$&$128.56 ^{+560.00 }_{-38.44 }$&$52.13 \pm0.03 $&$1,4$\\
061210&$5.31\pm0.47$&$0.41$&$-0.76 ^{+0.15 }_{-0.14 }$&$$&$544.04 ^{+763.39 }_{-309.56 }$&$51.24 \pm0.04 $&$2,4$\\
061222A&$8.53\pm0.26$&$2.088$&$-0.94 ^{+0.08 }_{-0.08 }$&$-2.41 ^{+0.17 }_{-0.73 }$&$283.00 ^{+36.00 }_{-25.00 }$&$52.93 \pm0.01 $&$2,3$\\
061222B&$1.59\pm0.36$&$3.355$&$-1.90 ^{+0.12 }_{-0.12 }$&$$&$52.00 ^{+43.55 }_{-26.01 }$&$52.48 \pm0.10 $&$1,3$\\
070103&$1.04\pm0.15$&$2.6208$&$-1.82 ^{+0.26 }_{-0.28 }$&$$&$57.69 ^{+370.47 }_{-43.13 }$&$51.98 \pm0.06 $&$1,4$\\
070110&$0.60\pm0.12$&$2.352$&$-1.15 ^{+0.44 }_{-0.40 }$&$$&$108.33 ^{+4977.94 }_{-33.17 }$&$51.51 \pm0.09 $&$1,4$\\
070129&$0.55\pm0.12$&$2.3384$&$-1.33 ^{+0.68 }_{-0.59 }$&$$&$47.68 ^{+63.55 }_{-12.48 }$&$51.38 \pm0.09 $&$1,4$\\
070208&$0.90\pm0.22$&$1.165$&$0.00 ^{+1.21 }_{-0.60 }$&$$&$66.00 ^{+108.27 }_{-19.96 }$&$50.76 \pm0.11 $&$1,3$\\
070223&$0.69\pm0.15$&$1.6295$&$-1.20 ^{+0.36 }_{-0.36 }$&$$&$61.00 ^{+20.57 }_{-9.07 }$&$51.09 \pm0.09 $&$1,3$\\
070224&$0.34\pm0.11$&$1.9922$&$-1.00 ^{*}$&$-2.27 ^{+0.28 }_{-0.32 }$&$29.58 ^{+12.80 }_{-28.05 }$&$51.10 \pm0.14 $&$1,4$\\
070306&$4.07\pm0.21$&$1.4959$&$-1.67 ^{+0.10 }_{-0.10 }$&$$&$119.60 ^{+535.79 }_{-38.67 }$&$51.96 \pm0.02 $&$1,4$\\
070318&$1.76\pm0.15$&$0.836$&$-1.41 ^{+0.08 }_{-0.08 }$&$$&$196.19 ^{+445.33 }_{-78.00 }$&$51.00 \pm0.04 $&$1,4$\\
070328&$4.22\pm0.24$&$2.0627$&$-1.14 ^{+0.04 }_{-0.04 }$&$$&$458.62 ^{+674.95 }_{-180.14 }$&$52.63 \pm0.02 $&$1,4$\\
070330&$0.88\pm0.14$&$2.4$&$-0.33 ^{+1.07 }_{-1.07 }$&$-2.20 ^{+0.40 }_{-0.40 }$&$36.00 ^{+4.00 }_{-4.00 }$&$51.72 \pm0.07 $&$3,3$\\
070411&$0.91\pm0.13$&$2.954$&$-1.70 ^{+0.10 }_{-0.10 }$&$$&$119.81 ^{+555.97 }_{-39.48 }$&$52.05 \pm0.06 $&$1,4$\\
070419A&$0.20\pm0.10$&$0.97$&$0.00 ^{+1.21 }_{-1.21 }$&$$&$27.00 ^{+9.68 }_{-11.49 }$&$49.80 \pm0.22 $&$2,3$\\
070419B&$1.38\pm0.17$&$1.9591$&$-1.36 ^{+0.19 }_{-0.18 }$&$$&$172.74 ^{+730.74 }_{-56.70 }$&$51.79 \pm0.05 $&$1,4$\\
070506&$0.96\pm0.13$&$2.31$&$5.00 ^{+0.00 }_{-3.18 }$&$-2.00 ^{+0.00 }_{-0.37 }$&$31.03 ^{+3.58 }_{-4.56 }$&$51.84 \pm0.06 $&$1,4$\\
070508&$24.10\pm0.61$&$0.82$&$-0.81 ^{+0.04 }_{-0.04 }$&$$&$188.00 ^{+4.84 }_{-4.84 }$&$52.15 \pm0.01 $&$3,3$\\
070518&$0.68\pm0.13$&$1.161$&$-1.00 ^{*}$&$-2.13 ^{+0.27 }_{-0.29 }$&$34.99 ^{+32.33 }_{-33.43 }$&$50.91 \pm0.08 $&$1,4$\\
070521&$6.53\pm0.27$&$2.0865$&$-0.93 ^{+0.07 }_{-0.07 }$&$$&$222.00 ^{+16.33 }_{-12.70 }$&$52.63 \pm0.02 $&$1,3$\\
070529&$1.43\pm0.36$&$2.4996$&$-1.30 ^{+0.12 }_{-0.12 }$&$$&$250.00 ^{+375.02 }_{-78.63 }$&$52.15 \pm0.11 $&$1,3$\\
070611&$0.82\pm0.21$&$2.04$&$0.00 ^{+1.21 }_{-0.60 }$&$$&$67.00 ^{+68.35 }_{-15.73 }$&$51.33 \pm0.11 $&$1,3$\\
070612A&$1.51\pm0.38$&$0.617$&$-1.58 ^{+0.06 }_{-0.06 }$&$$&$149.00 ^{+285.50 }_{-35.69 }$&$50.58 \pm0.11 $&$2,3$\\
070621&$2.50\pm0.30$&$1.5$&$-1.55 ^{+0.04 }_{-0.04 }$&$$&$157.00 ^{+230.45 }_{-30.24 }$&$51.76 \pm0.05 $&$3,3$\\
070714A&$2.70\pm0.20$&$2.58$&$-2.56 ^{+0.21 }_{-0.20 }$&$$&$250.00 ^{*}$&$53.09 \pm0.03 $&$3,3$\\
070714B&$2.70\pm0.20$&$0.92$&$-0.97 ^{+0.06 }_{-0.06 }$&$-2.12 ^{+0.42 }_{-7.88 }$&$543.75 ^{+355.73 }_{-178.13 }$&$51.84 \pm0.03 $&$1,3$\\
070721B&$1.50\pm0.30$&$3.626$&$-1.20 ^{+0.06 }_{-0.06 }$&$$&$410.00 ^{+489.94 }_{-120.97 }$&$52.71 \pm0.09 $&$1,3$\\
070802&$0.40\pm0.10$&$2.45$&$-1.82 ^{+0.27 }_{-0.29 }$&$$&$54.88 ^{+266.81 }_{-35.43 }$&$51.49 \pm0.11 $&$1,4$\\
070808&$2.00\pm0.20$&$1.35$&$-1.50 ^{+0.12 }_{-0.12 }$&$$&$135.00 ^{+270.37 }_{-28.43 }$&$51.52 \pm0.04 $&$3,3$\\
070810A&$1.90\pm0.20$&$2.17$&$-0.70 ^{+0.48 }_{-0.48 }$&$$&$41.00 ^{+4.84 }_{-4.84 }$&$51.72 \pm0.05 $&$2,3$\\
071003&$6.30\pm0.40$&$1.6043$&$-0.97 ^{+0.07 }_{-0.07 }$&$$&$799.00 ^{+124.00 }_{-100.00 }$&$52.82 \pm0.03 $&$1,3$\\
071010A&$0.80\pm0.30$&$0.98$&$-1.00 ^{*}$&$-2.11 ^{+0.39 }_{-0.44 }$&$36.79 ^{+49.23 }_{-35.22 }$&$50.81 \pm0.16 $&$2,4$\\
071010B&$7.70\pm0.30$&$0.947$&$-1.25 ^{+0.45 }_{-0.30 }$&$-2.65 ^{+0.18 }_{-0.30 }$&$52.00 ^{+6.05 }_{-8.47 }$&$51.62 \pm0.02 $&$1,3$\\
071011&$1.70\pm0.30$&$5$&$-1.40 ^{+0.06 }_{-0.06 }$&$$&$370.00 ^{+465.74 }_{-120.97 }$&$53.01 \pm0.08 $&$3,3$\\
071020&$8.40\pm0.30$&$2.145$&$-0.65 ^{+0.27 }_{-0.32 }$&$$&$322.00 ^{+80.00 }_{-53.00 }$&$52.98 \pm0.02 $&$1,3$\\
071021&$0.70\pm0.10$&$2.452$&$-1.73 ^{+0.22 }_{-0.23 }$&$$&$74.08 ^{+424.00 }_{-28.83 }$&$51.70 \pm0.06 $&$1,4$\\
071025&$1.60\pm0.20$&$5.2$&$-1.67 ^{+0.04 }_{-0.04 }$&$$&$165.00 ^{+301.83 }_{-35.69 }$&$52.91 \pm0.05 $&$1,3$\\
071028B&$1.40\pm0.50$&$0.94$&$$&$$&$$&$51.28 \pm0.16 $&$2,3$\\
071031&$0.50\pm0.10$&$2.692$&$-1.00 ^{*}$&$-2.27 ^{+0.29 }_{-0.32 }$&$12.25 ^{+6.19 }_{-11.25 }$&$51.65 \pm0.09 $&$1,4$\\
071101&$0.40\pm0.10$&$3.7$&$-1.10 ^{+0.40 }_{-0.40 }$&$-2.20 ^{+0.40 }_{-0.40 }$&$31.00 ^{+18.00 }_{-18.00 }$&$51.85 \pm0.11 $&$3,3$\\
071112C&$8.00\pm1.00$&$0.823$&$-1.09 ^{+0.04 }_{-0.04 }$&$$&$250.00 ^{*}$&$51.73 \pm0.05 $&$1,3$\\
071117&$11.30\pm0.40$&$1.331$&$-1.53 ^{+0.16 }_{-0.15 }$&$$&$278.00 ^{+236.00 }_{-79.00 }$&$52.38 \pm0.02 $&$1,3$\\
071122&$0.40\pm0.20$&$1.14$&$-1.60 ^{+0.24 }_{-0.24 }$&$$&$111.00 ^{+317.55 }_{-40.53 }$&$50.63 \pm0.22 $&$1,3$\\
080123&$1.80\pm0.40$&$0.495$&$-1.34 ^{+0.14 }_{-0.14 }$&$$&$250.00 ^{*}$&$50.52 \pm0.10 $&$3,3$\\
080129&$0.20\pm0.10$&$4.349$&$-1.20 ^{+0.12 }_{-0.12 }$&$$&$250.00 ^{+296.38 }_{-90.73 }$&$51.88 \pm0.22 $&$1,3$\\
080205&$1.40\pm0.20$&$2.72$&$-2.00 ^{+0.06 }_{-0.06 }$&$$&$50.00 ^{+21.17 }_{-29.03 }$&$52.32 \pm0.06 $&$1,3$\\
080207&$1.00\pm0.30$&$2.0858$&$-1.05 ^{+0.41 }_{-0.38 }$&$$&$102.50 ^{+127.50 }_{-23.41 }$&$51.58 \pm0.13 $&$1,4$\\
080210&$1.60\pm0.20$&$2.641$&$-1.75 ^{+0.12 }_{-0.12 }$&$$&$90.48 ^{+472.05 }_{-36.46 }$&$52.17 \pm0.05 $&$1,4$\\
080310&$1.30\pm0.20$&$2.42$&$-1.00 ^{*}$&$-2.35 ^{+0.17 }_{-0.17 }$&$22.31 ^{+8.32 }_{-20.83 }$&$51.87 \pm0.07 $&$1,4$\\
080319A&$1.20\pm0.20$&$2.0265$&$-1.60 ^{+0.06 }_{-0.06 }$&$$&$105.00 ^{+181.46 }_{-21.17 }$&$51.72 \pm0.07 $&$1,3$\\
080319B&$24.80\pm0.50$&$0.937$&$-0.82 ^{+0.01 }_{-0.01 }$&$-3.87 ^{+0.44 }_{-1.09 }$&$651.00 ^{+13.00 }_{-14.00 }$&$52.84 \pm0.01 $&$1,3$\\
080319C&$5.20\pm0.30$&$1.95$&$-1.00 ^{+0.18 }_{-0.18 }$&$$&$157.00 ^{+183.27 }_{-30.24 }$&$52.34 \pm0.03 $&$1,3$\\
080320&$0.60\pm0.10$&$7$&$-1.60 ^{+0.18 }_{-0.18 }$&$$&$95.00 ^{+293.36 }_{-23.59 }$&$52.70 \pm0.07 $&$3,3$\\
080325&$1.40\pm0.60$&$1.78$&$-1.67 ^{+0.17 }_{-0.18 }$&$$&$110.52 ^{+398.86 }_{-41.46 }$&$51.67 \pm0.19 $&$1,4$\\
080330&$0.90\pm0.20$&$1.51$&$-1.00 ^{*}$&$-2.38 ^{+0.41 }_{-0.48 }$&$20.19 ^{+6.58 }_{-19.19 }$&$51.21 \pm0.10 $&$1,4$\\
080411&$43.20\pm0.90$&$1.03$&$-1.68 ^{+0.03 }_{-0.03 }$&$$&$405.52 ^{+806.07 }_{-185.53 }$&$52.74 \pm0.01 $&$1,4$\\
080413A&$5.60\pm0.20$&$2.433$&$-1.56 ^{+0.04 }_{-0.04 }$&$$&$149.00 ^{+223.80 }_{-25.40 }$&$52.62 \pm0.02 $&$2,3$\\
080413B&$18.70\pm0.80$&$1.1$&$-1.26 ^{+0.16 }_{-0.16 }$&$$&$73.30 ^{+9.56 }_{-9.56 }$&$52.13 \pm0.02 $&$1,3$\\
080430&$2.60\pm0.20$&$0.767$&$-1.74 ^{+0.09 }_{-0.09 }$&$$&$122.14 ^{+553.65 }_{-46.59 }$&$51.06 \pm0.03 $&$1,4$\\
080515&$3.90\pm0.70$&$2.47$&$0.00 ^{+0.60 }_{-0.60 }$&$$&$26.00 ^{+3.02 }_{-3.02 }$&$52.10 \pm0.08 $&$1,3$\\
080516&$1.80\pm0.30$&$3.2$&$-1.80 ^{+0.18 }_{-0.18 }$&$$&$67.00 ^{+243.16 }_{-22.98 }$&$52.43 \pm0.07 $&$2,3$\\
080520&$0.50\pm0.10$&$1.545$&$-1.00 ^{*}$&$-3.35 ^{+0.63 }_{-0.88 }$&$11.08 ^{+3.06 }_{-10.08 }$&$51.02 \pm0.09 $&$1,4$\\
080602&$2.90\pm0.20$&$1.8204$&$-1.51 ^{+0.13 }_{-0.14 }$&$$&$116.18 ^{+251.68 }_{-43.94 }$&$51.98 \pm0.03 $&$1,4$\\
080603B&$3.50\pm0.20$&$2.69$&$-1.23 ^{+0.45 }_{-0.33 }$&$$&$102.00 ^{+71.98 }_{-16.94 }$&$52.41 \pm0.02 $&$1,3$\\
080604&$0.40\pm0.10$&$1.416$&$-1.70 ^{+0.12 }_{-0.12 }$&$$&$74.00 ^{+256.46 }_{-17.54 }$&$50.85 \pm0.11 $&$1,3$\\
080605&$19.90\pm0.60$&$1.6398$&$-1.13 ^{+0.15 }_{-0.15 }$&$$&$227.84 ^{+249.64 }_{-64.41 }$&$52.84 \pm0.01 $&$1,4$\\
080607&$23.10\pm1.10$&$3.036$&$-1.17 ^{+0.04 }_{-0.04 }$&$$&$902.50 ^{+1169.78 }_{-460.32 }$&$53.95 \pm0.02 $&$1,4$\\
080707&$1.00\pm0.10$&$1.23$&$-1.71 ^{+0.19 }_{-0.19 }$&$$&$74.08 ^{+317.31 }_{-31.97 }$&$51.10 \pm0.04 $&$1,4$\\
080710&$1.00\pm0.20$&$0.845$&$-1.30 ^{+0.12 }_{-0.12 }$&$$&$300.00 ^{+332.67 }_{-120.97 }$&$50.88 \pm0.09 $&$1,3$\\
080721&$20.90\pm1.80$&$2.591$&$-0.95 ^{+0.09 }_{-0.09 }$&$$&$604.96 ^{+979.46 }_{-292.44 }$&$53.74 \pm0.04 $&$1,4$\\
080804&$3.10\pm0.40$&$2.2045$&$-1.03 ^{+0.11 }_{-0.11 }$&$$&$405.52 ^{+650.03 }_{-204.15 }$&$52.56 \pm0.06 $&$1,4$\\
080805&$1.10\pm0.10$&$1.505$&$-1.54 ^{+0.09 }_{-0.09 }$&$$&$300.42 ^{+552.02 }_{-188.93 }$&$51.51 \pm0.04 $&$1,4$\\
080810&$2.00\pm0.20$&$3.35$&$-1.31 ^{+0.12 }_{-0.12 }$&$$&$366.93 ^{+616.00 }_{-217.69 }$&$52.69 \pm0.04 $&$1,4$\\
080905B&$0.50\pm0.10$&$2.374$&$-0.85 ^{+0.14 }_{-0.14 }$&$-2.28 ^{+0.38 }_{-0.38 }$&$179.00 ^{+36.29 }_{-36.29 }$&$51.74 \pm0.09 $&$1,3$\\
080906&$1.00\pm0.20$&$2.1$&$-1.58 ^{+0.05 }_{-0.05 }$&$$&$105.00 ^{+142.75 }_{-18.75 }$&$51.67 \pm0.09 $&$1,3$\\
080913&$1.40\pm0.20$&$6.695$&$-0.40 ^{+0.88 }_{-0.74 }$&$$&$98.20 ^{+135.40 }_{-22.89 }$&$52.91 \pm0.06 $&$1,4$\\
080916A&$2.70\pm0.20$&$0.689$&$-1.00 ^{+0.23 }_{-0.19 }$&$$&$129.00 ^{+19.96 }_{-12.70 }$&$50.88 \pm0.03 $&$2,3$\\
080916B&$0.60\pm0.20$&$4.35$&$-1.50 ^{+0.12 }_{-0.12 }$&$$&$100.00 ^{+228.03 }_{-22.38 }$&$52.19 \pm0.14 $&$3,3$\\
080928&$2.10\pm0.10$&$1.692$&$-1.70 ^{+0.06 }_{-0.06 }$&$$&$74.00 ^{+146.98 }_{-15.73 }$&$51.77 \pm0.02 $&$1,3$\\
081007&$2.60\pm0.40$&$0.5295$&$-1.35 ^{+1.02 }_{-0.64 }$&$$&$27.25 ^{+11.37 }_{-15.73 }$&$50.45 \pm0.07 $&$1,4$\\
081008&$1.30\pm0.10$&$1.9685$&$-1.30 ^{+0.34 }_{-0.31 }$&$$&$105.25 ^{+426.59 }_{-29.00 }$&$51.66 \pm0.03 $&$1,4$\\
081028A&$0.50\pm0.10$&$3.038$&$-1.30 ^{+0.24 }_{-0.24 }$&$$&$67.00 ^{+18.75 }_{-7.86 }$&$51.64 \pm0.09 $&$2,3$\\
081029&$0.50\pm0.20$&$3.8479$&$-1.40 ^{+0.12 }_{-0.12 }$&$$&$300.00 ^{+332.67 }_{-120.97 }$&$52.17 \pm0.17 $&$1,3$\\
081102A&$1.40\pm0.30$&$3.04$&$-0.84 ^{+0.15 }_{-0.17 }$&$$&$80.00 ^{+9.68 }_{-6.05 }$&$52.06 \pm0.09 $&$3,3$\\
081109A&$1.10\pm0.00$&$0.9787$&$-0.67 ^{+1.85 }_{-1.85 }$&$-1.65 ^{+0.03 }_{-0.03 }$&$32.40 ^{+38.30 }_{-38.30 }$&$51.34 \pm0.00 $&$1,3$\\
081118&$0.60\pm0.20$&$2.58$&$-1.00 ^{*}$&$-2.14 ^{+0.18 }_{-0.19 }$&$36.79 ^{+12.18 }_{-35.22 }$&$51.68 \pm0.14 $&$1,4$\\
081121&$4.40\pm1.00$&$2.512$&$-0.52 ^{+0.56 }_{-0.50 }$&$$&$133.77 ^{+163.06 }_{-33.28 }$&$52.50 \pm0.10 $&$1,4$\\
081203A&$2.90\pm0.20$&$2.1$&$-1.44 ^{+0.04 }_{-0.04 }$&$$&$201.00 ^{+266.14 }_{-45.36 }$&$52.22 \pm0.03 $&$2,3$\\
081210&$2.50\pm0.20$&$2.0631$&$-1.40 ^{+0.06 }_{-0.06 }$&$$&$220.00 ^{+356.87 }_{-66.53 }$&$52.15 \pm0.03 $&$1,3$\\
081221&$18.20\pm0.50$&$2.26$&$-1.10 ^{+0.15 }_{-0.14 }$&$-29.44 ^{+27.20 }_{-85.56 }$&$82.60 ^{+7.67 }_{-5.40 }$&$52.89 \pm0.01 $&$1,4$\\
081222&$7.70\pm0.20$&$2.77$&$-1.04 ^{+0.17 }_{-0.17 }$&$$&$136.22 ^{+53.27 }_{-24.01 }$&$52.84 \pm0.01 $&$1,4$\\
081228&$0.60\pm0.10$&$3.44$&$-2.00 ^{+0.18 }_{-0.18 }$&$$&$39.00 ^{+47.18 }_{-20.57 }$&$52.20 \pm0.07 $&$1,3$\\
081230&$0.70\pm0.10$&$2$&$-1.00 ^{+0.48 }_{-0.48 }$&$$&$52.00 ^{+11.49 }_{-8.47 }$&$51.26 \pm0.06 $&$1,3$\\
090102&$5.50\pm0.80$&$1.547$&$-1.36 ^{+0.10 }_{-0.10 }$&$$&$366.93 ^{+616.00 }_{-217.69 }$&$52.31 \pm0.06 $&$1,4$\\
090113&$2.50\pm0.20$&$1.7493$&$-1.52 ^{+0.10 }_{-0.10 }$&$$&$128.40 ^{+434.15 }_{-39.65 }$&$51.89 \pm0.03 $&$1,4$\\
090201&$14.70\pm1.00$&$2.1$&$-0.97 ^{+0.05 }_{-0.05 }$&$-2.80 ^{+0.31 }_{-0.31 }$&$254.84 $&$53.08 \pm0.03 $&$1,3$\\
090205&$0.50\pm0.10$&$4.6497$&$-0.89 ^{+1.69 }_{-1.10 }$&$$&$33.29 ^{+31.16 }_{-28.99 }$&$51.95 \pm0.09 $&$1,4$\\
090313&$0.80\pm0.30$&$3.375$&$-1.89 ^{+0.29 }_{-0.32 }$&$$&$54.88 ^{+202.37 }_{-51.09 }$&$52.18 \pm0.16 $&$1,4$\\
090401B&$23.10\pm0.50$&$3.1$&$-0.89 ^{+0.11 }_{-0.11 }$&$-2.32 ^{+0.21 }_{-0.39 }$&$205.00 ^{+55.00 }_{-37.00 }$&$53.69 \pm0.01 $&$3,3$\\
090404&$1.90\pm0.20$&$3$&$-1.80 ^{+0.24 }_{-0.12 }$&$$&$31.00 ^{+2.42 }_{-18.15 }$&$52.35 \pm0.05 $&$1,3$\\
090407&$0.60\pm0.10$&$1.4485$&$-1.48 ^{+0.37 }_{-0.38 }$&$$&$100.00 ^{+497.93 }_{-49.96 }$&$51.02 \pm0.07 $&$1,4$\\
090417B&$0.30\pm0.10$&$0.345$&$-1.70 ^{+0.06 }_{-0.06 }$&$$&$111.00 ^{+428.24 }_{-24.80 }$&$49.26 \pm0.14 $&$1,3$\\
090418A&$1.90\pm0.30$&$1.608$&$-1.30 ^{+0.09 }_{-0.09 }$&$$&$610.00 ^{+554.00 }_{-215.00 }$&$52.04 \pm0.07 $&$2,3$\\
090423&$1.70\pm0.20$&$8.26$&$-0.70 ^{+0.62 }_{-0.55 }$&$$&$50.55 ^{+9.35 }_{-5.66 }$&$53.07 \pm0.05 $&$1,4$\\
090424&$71.00\pm2.00$&$0.544$&$-1.08 ^{+0.16 }_{-0.15 }$&$$&$126.08 ^{+37.02 }_{-18.86 }$&$52.05 \pm0.01 $&$1,4$\\
090429B&$1.60\pm0.20$&$9.2$&$-0.47 ^{+0.47 }_{-0.47 }$&$$&$42.10 ^{+3.39 }_{-3.39 }$&$53.10 \pm0.05 $&$1,3$\\
090516A&$1.60\pm0.20$&$4.109$&$-1.52 ^{+0.03 }_{-0.03 }$&$-2.30 ^{+0.16 }_{-0.16 }$&$142.00 ^{+15.73 }_{-15.73 }$&$52.67 \pm0.05 $&$2,3$\\
090519&$0.60\pm0.20$&$3.85$&$-0.98 ^{+0.22 }_{-0.21 }$&$$&$405.52 ^{+806.07 }_{-223.52 }$&$52.44 \pm0.14 $&$1,4$\\
090529&$0.40\pm0.10$&$2.625$&$-1.00 ^{+0.60 }_{-0.60 }$&$$&$43.00 ^{+18.75 }_{-10.28 }$&$51.29 \pm0.11 $&$1,3$\\
090530&$2.50\pm0.30$&$1.266$&$-1.60 ^{+0.12 }_{-0.12 }$&$$&$90.00 ^{+130.05 }_{-19.96 }$&$51.51 \pm0.05 $&$1,3$\\
090618&$38.90\pm0.80$&$0.54$&$-1.26 ^{+0.04 }_{-0.01 }$&$-2.50 ^{+0.09 }_{-0.20 }$&$155.50 ^{+6.71 }_{-6.35 }$&$51.93 \pm0.01 $&$1,3$\\
090709A&$7.80\pm0.30$&$1.8$&$-0.85 ^{+0.05 }_{-0.05 }$&$-2.70 ^{+0.15 }_{-0.15 }$&$478.93 $&$52.93 \pm0.02 $&$3,3$\\
090715B&$3.80\pm0.20$&$3$&$-1.10 ^{+0.40 }_{-0.34 }$&$$&$134.00 ^{+56.00 }_{-30.00 }$&$52.61 \pm0.02 $&$1,3$\\
090726&$0.70\pm0.20$&$2.71$&$-1.20 ^{+0.79 }_{-0.48 }$&$$&$27.00 ^{+7.86 }_{-13.31 }$&$51.60 \pm0.12 $&$1,3$\\
090809&$1.10\pm0.20$&$2.737$&$-1.70 ^{+0.51 }_{-0.57 }$&$$&$74.08 ^{+825.80 }_{-68.05 }$&$52.00 \pm0.08 $&$1,4$\\
090812&$3.60\pm0.20$&$2.452$&$-1.29 ^{+0.06 }_{-0.06 }$&$$&$271.83 ^{+522.06 }_{-117.45 }$&$52.55 \pm0.02 $&$1,4$\\
090814A&$0.60\pm0.20$&$0.7$&$-1.84 ^{+0.19 }_{-0.18 }$&$$&$250.00 ^{*}$&$50.43 \pm0.14 $&$3,3$\\
090904B&$1.90\pm0.20$&$5$&$-1.13 ^{+0.04 }_{-0.04 }$&$-2.30 ^{+0.07 }_{-0.07 }$&$113.00 ^{+6.65 }_{-6.65 }$&$52.91 \pm0.05 $&$3,3$\\
090926B&$3.20\pm0.30$&$1.24$&$-0.13 ^{+0.04 }_{-0.04 }$&$$&$91.00 ^{+1.21 }_{-1.21 }$&$51.48 \pm0.04 $&$1,3$\\
090927&$2.00\pm0.20$&$1.37$&$-1.77 ^{+0.34 }_{-0.37 }$&$$&$60.65 ^{+527.35 }_{-55.11 }$&$51.53 \pm0.04 $&$1,4$\\
091018&$10.30\pm0.40$&$0.971$&$-1.69 ^{+0.32 }_{-0.30 }$&$$&$22.74 ^{+8.45 }_{-21.57 }$&$51.80 \pm0.02 $&$1,4$\\
091020&$4.20\pm0.30$&$1.71$&$-1.24 ^{+0.31 }_{-0.29 }$&$$&$140.04 ^{+1827.70 }_{-45.38 }$&$52.08 \pm0.03 $&$1,4$\\
091024&$2.00\pm0.30$&$1.092$&$-1.28 ^{+0.07 }_{-0.07 }$&$$&$349.03 ^{+599.50 }_{-172.23 }$&$51.50 \pm0.07 $&$1,4$\\
091029&$1.80\pm0.10$&$2.752$&$-1.40 ^{+0.32 }_{-0.30 }$&$$&$59.35 ^{+19.53 }_{-8.01 }$&$52.10 \pm0.02 $&$1,4$\\
091109A&$1.30\pm0.40$&$3.076$&$-1.23 ^{+0.27 }_{-0.26 }$&$$&$448.17 ^{+847.42 }_{-272.07 }$&$52.49 \pm0.13 $&$2,4$\\
091127&$46.50\pm2.70$&$0.49$&$-1.00 ^{*}$&$-2.13 ^{+0.08 }_{-0.09 }$&$27.25 ^{+5.23 }_{-26.25 }$&$51.83 \pm0.03 $&$1,4$\\
091208B&$15.20\pm1.00$&$1.063$&$-1.29 ^{+0.08 }_{-0.07 }$&$-2.28 ^{+0.07 }_{-0.08 }$&$108.63 ^{+73.34 }_{-57.60 }$&$52.22 \pm0.03 $&$1,3$\\
100219A&$0.40\pm0.10$&$4.6667$&$-1.37 ^{+0.26 }_{-0.26 }$&$$&$122.14 ^{+422.09 }_{-51.71 }$&$52.09 \pm0.11 $&$1,4$\\
100302A&$0.50\pm0.10$&$4.813$&$-1.00 ^{*}$&$-2.03 ^{+0.23 }_{-0.24 }$&$42.74 ^{+31.28 }_{-41.13 }$&$52.27 \pm0.09 $&$1,4$\\
100316A&$2.30\pm0.40$&$3.155$&$-1.54 ^{+0.23 }_{-0.23 }$&$$&$128.40 ^{+434.15 }_{-60.53 }$&$52.48 \pm0.08 $&$1,4$\\
100316B&$1.30\pm0.10$&$1.18$&$-1.00 ^{*}$&$-2.29 ^{+0.22 }_{-0.24 }$&$28.65 ^{+13.90 }_{-27.13 }$&$51.13 \pm0.03 $&$1,4$\\
100316D&$0.10\pm0.00$&$0.059$&$-2.29 ^{+0.41 }_{-0.41 }$&$$&$250.00 ^{*}$&$49.39 \pm0.00 $&$3,3$\\
100413A&$0.70\pm0.10$&$3.9$&$-1.22 ^{+0.03 }_{-0.03 }$&$-2.20 $&$959.00 ^{+200.21 }_{-200.21 }$&$52.66 \pm0.06 $&$2,3$\\
100418A&$1.00\pm0.20$&$0.6235$&$-1.00 ^{*}$&$-2.06 ^{+0.27 }_{-0.29 }$&$36.79 ^{+16.80 }_{-35.79 }$&$50.47 \pm0.09 $&$1,4$\\
100424A&$0.40\pm0.10$&$2.465$&$-1.87 ^{+0.14 }_{-0.14 }$&$$&$54.88 ^{+98.76 }_{-32.70 }$&$51.52 \pm0.11 $&$1,4$\\
100425A&$1.40\pm0.20$&$1.755$&$-0.53 ^{+2.83 }_{-1.46 }$&$$&$23.46 ^{+11.94 }_{-21.36 }$&$51.35 \pm0.06 $&$1,4$\\
100508A&$0.40\pm0.20$&$0.5201$&$-1.23 ^{+0.25 }_{-0.25 }$&$$&$201.38 ^{+557.34 }_{-101.31 }$&$49.87 \pm0.22 $&$1,4$\\
100513A&$0.60\pm0.10$&$4.772$&$-1.65 ^{+0.14 }_{-0.14 }$&$$&$81.87 ^{+258.16 }_{-23.28 }$&$52.31 \pm0.07 $&$1,4$\\
100606A&$1.60\pm0.40$&$1.5545$&$-1.05 ^{+0.14 }_{-0.14 }$&$$&$945.00 ^{+551.00 }_{-266.00 }$&$52.20 \pm0.11 $&$3,3$\\
100615A&$5.40\pm0.20$&$1.398$&$-0.90 ^{+0.18 }_{-0.18 }$&$-1.80 ^{+0.03 }_{-0.03 }$&$53.30 ^{+9.90 }_{-9.90 }$&$52.26 \pm0.02 $&$1,3$\\
100621A&$12.80\pm0.30$&$0.542$&$-1.70 ^{+0.08 }_{-0.08 }$&$-2.45 ^{+1.44 }_{-1.44 }$&$95.00 ^{+8.00 }_{-11.00 }$&$51.41 \pm0.01 $&$1,3$\\
100704A&$4.30\pm0.20$&$3.6$&$-0.93 ^{+0.05 }_{-0.05 }$&$-2.00 ^{+0.08 }_{-0.08 }$&$175.00 ^{+18.15 }_{-18.15 }$&$53.14 \pm0.02 $&$2,3$\\
100724A&$1.90\pm0.20$&$1.288$&$-0.76 ^{+0.01 }_{-0.01 }$&$-2.03 ^{+0.02 }_{-0.02 }$&$413.10 ^{+8.90 }_{-8.90 }$&$52.07 \pm0.05 $&$3,3$\\
100728A&$5.10\pm0.20$&$1.567$&$-0.75 ^{+0.01 }_{-0.01 }$&$-3.04 ^{+0.34 }_{-0.34 }$&$344.00 ^{+5.44 }_{-5.44 }$&$52.48 \pm0.02 $&$1,3$\\
100728B&$3.50\pm0.50$&$2.106$&$-0.80 ^{+0.12 }_{-0.12 }$&$-2.20 ^{+0.12 }_{-0.12 }$&$104.00 ^{+8.47 }_{-8.47 }$&$52.35 \pm0.06 $&$1,3$\\
100805A&$0.70\pm0.20$&$1.3$&$-1.00 ^{*}$&$$&$64.00 $&$50.83 \pm0.12 $&$2,3$\\
100814A&$2.50\pm0.20$&$1.44$&$-0.64 ^{+0.08 }_{-0.08 }$&$-2.02 ^{+0.07 }_{-0.07 }$&$106.00 ^{+8.47 }_{-8.47 }$&$51.92 \pm0.03 $&$1,3$\\
100816A&$10.90\pm0.40$&$0.8034$&$-1.00 ^{+0.24 }_{-0.18 }$&$$&$148.00 ^{+24.80 }_{-15.73 }$&$51.69 \pm0.02 $&$2,3$\\
100901A&$0.80\pm0.20$&$1.408$&$-1.55 ^{+0.22 }_{-0.23 }$&$$&$128.40 ^{+568.43 }_{-54.79 }$&$51.16 \pm0.11 $&$1,4$\\
100902A&$1.00\pm0.10$&$4.5$&$-1.96 ^{+0.14 }_{-0.14 }$&$$&$49.66 ^{+50.27 }_{-48.02 }$&$52.64 \pm0.04 $&$2,4$\\
100905A&$0.60\pm0.10$&$7.9$&$-1.27 ^{+0.22 }_{-0.21 }$&$$&$164.87 ^{+499.25 }_{-73.13 }$&$52.85 \pm0.07 $&$1,4$\\
100906A&$10.10\pm0.40$&$1.727$&$-1.34 ^{+0.05 }_{-0.05 }$&$-1.98 ^{+0.04 }_{-0.04 }$&$106.00 ^{+10.89 }_{-10.89 }$&$52.66 \pm0.02 $&$1,3$\\
101213A&$2.20\pm0.40$&$0.414$&$-1.37 ^{+0.09 }_{-0.09 }$&$$&$405.52 ^{+650.03 }_{-243.39 }$&$50.53 \pm0.08 $&$1,4$\\
101219B&$0.60\pm0.30$&$0.55$&$-1.37 ^{+0.72 }_{-0.72 }$&$-2.26 ^{+0.14 }_{-0.14 }$&$56.40 ^{+6.60 }_{-6.60 }$&$50.06 \pm0.22 $&$1,3$\\
110106B&$2.10\pm0.30$&$0.618$&$-1.40 ^{+0.11 }_{-0.12 }$&$$&$111.00 ^{+35.69 }_{-15.73 }$&$50.65 \pm0.06 $&$1,3$\\
110128A&$0.80\pm0.20$&$2.339$&$-1.26 ^{+0.25 }_{-0.25 }$&$$&$192.00 ^{+112.00 }_{-112.00 }$&$51.76 \pm0.11 $&$1,3$\\
110205A&$3.60\pm0.20$&$2.22$&$-1.52 ^{+0.09 }_{-0.09 }$&$-2.36 ^{+0.31 }_{-0.31 }$&$222.00 ^{+46.00 }_{-46.00 }$&$52.44 \pm0.02 $&$1,3$\\
110213A&$1.60\pm0.60$&$1.46$&$-1.44 ^{+0.03 }_{-0.03 }$&$-2.36 ^{+0.31 }_{-0.31 }$&$98.00 ^{+4.00 }_{-5.00 }$&$51.55 \pm0.16 $&$1,3$\\
110422A&$30.70\pm1.00$&$1.77$&$-0.65 ^{+0.04 }_{-0.04 }$&$-2.96 ^{+0.08 }_{-0.11 }$&$152.00 ^{+3.02 }_{-3.02 }$&$53.08 \pm0.01 $&$1,3$\\
110503A&$1.35\pm0.06$&$1.613$&$-0.88 ^{+0.15 }_{-0.15 }$&$$&$133.00 ^{+32.66 }_{-32.66 }$&$51.50 \pm0.02 $&$1,3$\\
110709B&$3.40\pm0.30$&$2.09£¿$&$-1.00 ^{+0.14 }_{-0.13 }$&$$&$278.00 ^{+43.00 }_{-32.00 }$&$52.41 \pm0.04 $&$1,3$\\
110715A&$53.90\pm1.10$&$0.82$&$-1.23 ^{+0.05 }_{-0.05 }$&$-2.70 ^{+0.12 }_{-0.30 }$&$120.00 ^{+7.26 }_{-6.65 }$&$52.43 \pm0.01 $&$1,3$\\
110726A&$1.00\pm0.20$&$1.04$&$-0.64 ^{+0.53 }_{-0.53 }$&$$&$46.50 ^{+7.14 }_{-7.14 }$&$50.66 \pm0.09 $&$3,3$\\
110731A&$11.00\pm0.30$&$2.83$&$-0.82 ^{+0.03 }_{-0.03 }$&$-2.32 ^{+0.02 }_{-0.03 }$&$311.52 ^{+35.46 }_{-31.52 }$&$53.43 \pm0.01 $&$1,3$\\
110801A&$1.10\pm0.20$&$1.858$&$-1.70 ^{+0.22 }_{-0.15 }$&$$&$140.00 ^{+1270.00 }_{-50.00 }$&$51.65 \pm0.08 $&$1,3$\\
110808A&$0.40\pm0.20$&$1.348$&$-1.13 ^{+0.08 }_{-0.08 }$&$$&$2960.00 ^{+1217.00 }_{-796.00 }$&$51.70 \pm0.22 $&$1,3$\\
110818A&$1.60\pm0.30$&$3.36$&$-1.11 ^{+0.17 }_{-0.17 }$&$-1.76 ^{+0.10 }_{-0.10 }$&$183.00 ^{+85.00 }_{-85.00 }$&$52.71 \pm0.08 $&$1,3$\\
111008A&$6.40\pm0.70$&$4.9898$&$-1.36 ^{+0.24 }_{-0.21 }$&$$&$149.00 ^{+52.00 }_{-28.00 }$&$53.40 \pm0.05 $&$1,3$\\
111107A&$1.20\pm0.20$&$2.893$&$-1.38 ^{+0.21 }_{-0.21 }$&$$&$27.74 ^{+8.22 }_{-8.22 }$&$51.95 \pm0.07 $&$1,3$\\
111123A&$0.90\pm0.10$&$3.1516$&$-1.30 ^{+0.26 }_{-0.24 }$&$$&$25.96 ^{+30.20 }_{-6.03 }$&$51.90 \pm0.05 $&$1,3$\\
111129A&$0.90\pm0.20$&$1.0796$&$-1.00 ^{*}$&$-2.65 ^{+0.44 }_{-0.55 }$&$16.53 ^{+4.83 }_{-15.53 }$&$50.82 \pm0.10 $&$1,4$\\
111209A&$0.50\pm0.10$&$0.677$&$-1.31 ^{+0.09 }_{-0.09 }$&$$&$310.00 ^{+53.00 }_{-53.00 }$&$50.35 \pm0.09 $&$2,3$\\
111212A&$0.80\pm0.30$&$0.269$&$$&$$&$$&$49.75 \pm0.16 $&$2,$\\
111215A&$0.50\pm0.20$&$2.06$&$-1.41 ^{+0.35 }_{-0.36 }$&$$&$122.14 ^{+553.65 }_{-57.14 }$&$51.34 \pm0.17 $&$1,4$\\
111225A&$0.70\pm0.10$&$0.297$&$-1.67 ^{+0.16 }_{-0.16 }$&$$&$77.88 ^{+250.45 }_{-27.47 }$&$49.43 \pm0.06 $&$1,4$\\
111228A&$12.40\pm0.50$&$0.714$&$-1.90 ^{+0.06 }_{-0.06 }$&$-2.70 ^{+0.18 }_{-0.18 }$&$34.00 ^{+1.81 }_{-1.81 }$&$51.68 \pm0.02 $&$1,3$\\
111229A&$1.00\pm0.20$&$1.3805$&$-1.00 ^{*}$&$-2.01 ^{+0.31 }_{-0.35 }$&$44.93 ^{+56.78 }_{-43.31 }$&$51.34 \pm0.09 $&$1,4$\\
120118B&$2.20\pm0.30$&$2.943$&$-0.30 ^{+0.29 }_{-0.29 }$&$-2.55 ^{+0.15 }_{-0.15 }$&$43.10 ^{+3.90 }_{-3.90 }$&$52.22 \pm0.06 $&$1,3$\\
120119A&$10.30\pm0.30$&$1.728$&$-0.96 ^{+0.03 }_{-0.03 }$&$-2.37 ^{+0.09 }_{-0.09 }$&$183.00 ^{+7.00 }_{-7.00 }$&$52.69 \pm0.01 $&$1,3$\\
120211A&$0.50\pm0.20$&$2.4$&$-1.44 ^{+0.26 }_{-0.27 }$&$$&$116.18 ^{+410.33 }_{-53.94 }$&$51.50 \pm0.17 $&$1,4$\\
120224A&$0.90\pm0.20$&$1.1$&$-1.00 ^{*}$&$-2.22 ^{+0.44 }_{-0.54 }$&$33.29 ^{+21.80 }_{-32.29 }$&$50.92 \pm0.10 $&$1,4$\\
120326A&$4.60\pm0.20$&$1.798$&$-0.98 ^{+0.08 }_{-0.08 }$&$-2.53 ^{+0.09 }_{-0.09 }$&$46.50 ^{+2.24 }_{-2.24 }$&$52.07 \pm0.02 $&$1,3$\\
120327A&$3.90\pm0.20$&$2.813$&$-1.14 ^{+0.26 }_{-0.28 }$&$$&$27.82 ^{+5.97 }_{-21.01 }$&$52.37 \pm0.02 $&$1,3$\\
120404A&$1.20\pm0.20$&$2.876$&$-1.84 ^{+0.14 }_{-0.14 }$&$$&$60.65 ^{+167.03 }_{-25.27 }$&$52.15 \pm0.07 $&$1,4$\\
120422A&$0.60\pm0.20$&$0.283$&$-1.90 ^{+0.44 }_{-0.49 }$&$$&$40.66 ^{+306.42 }_{-38.26 }$&$49.38 \pm0.14 $&$1,4$\\
120521C&$1.90\pm0.20$&$6$&$-1.68 ^{+0.12 }_{-0.12 }$&$$&$77.88 ^{+195.76 }_{-24.59 }$&$53.05 \pm0.05 $&$1,4$\\
120712A&$2.40\pm0.20$&$4.1745$&$-0.60 ^{+0.20 }_{-0.20 }$&$-1.80 ^{+0.20 }_{-0.20 }$&$23.96 ^{+5.02 }_{-5.02 }$&$52.94 \pm0.04 $&$1,3$\\
120714B&$0.40\pm0.10$&$0.3984$&$-0.29 ^{+0.96 }_{-0.80 }$&$$&$43.48 ^{+7.31 }_{-18.54 }$&$49.21 \pm0.11 $&$1,3$\\
120722A&$1.00\pm0.30$&$0.9586$&$-1.95 ^{+0.26 }_{-0.28 }$&$$&$49.66 ^{+50.27 }_{-48.66 }$&$50.95 \pm0.13 $&$1,4$\\
120724A&$0.60\pm0.20$&$1.48$&$-0.53 ^{+0.93 }_{-0.93 }$&$$&$27.60 ^{+4.54 }_{-4.54 }$&$50.78 \pm0.14 $&$1,3$\\
120729A&$2.90\pm0.20$&$0.8$&$-1.62 ^{+0.08 }_{-0.08 }$&$$&$106.67 $&$51.11 \pm0.03 $&$1,3$\\
120802A&$3.00\pm0.20$&$3.796$&$-1.21 ^{+0.28 }_{-0.28 }$&$$&$57.20 ^{+11.73 }_{-11.73 }$&$52.61 \pm0.03 $&$1,3$\\
120805A&$0.37\pm0.20$&$3.1$&$-0.08 ^{+1.77 }_{-1.27 }$&$$&$84.33 ^{+764.14 }_{-21.60 }$&$51.49 \pm0.23 $&$1,4$\\
120811C&$4.10\pm0.20$&$2.671$&$-1.40 ^{+0.18 }_{-0.18 }$&$$&$42.90 ^{+3.45 }_{-3.45 }$&$52.41 \pm0.02 $&$1,3$\\
120815A&$2.20\pm0.30$&$2.358$&$-1.00 ^{*}$&$-2.47 ^{+0.26 }_{-0.30 }$&$23.46 ^{+10.13 }_{-21.36 }$&$52.04 \pm0.06 $&$1,4$\\
120907A&$2.90\pm0.40$&$0.97$&$-0.75 ^{+0.25 }_{-0.25 }$&$$&$154.52 ^{+32.89 }_{-32.89 }$&$51.35 \pm0.06 $&$1,3$\\
120909A&$1.80\pm0.30$&$3.93$&$-0.83 ^{+0.08 }_{-0.08 }$&$-1.92 ^{+0.09 }_{-0.09 }$&$195.00 ^{+25.00 }_{-25.00 }$&$52.92 \pm0.07 $&$2,3$\\
120922A&$2.00\pm0.20$&$3.1$&$-1.60 ^{+0.42 }_{-0.42 }$&$-2.30 ^{+0.06 }_{-0.06 }$&$37.70 ^{+2.12 }_{-2.12 }$&$52.41 \pm0.04 $&$1,3$\\
120923A&$0.60\pm0.10$&$7.8$&$-0.29 ^{+1.00 }_{-1.00 }$&$$&$44.40 ^{+6.41 }_{-6.41 }$&$52.51 \pm0.07 $&$1,3$\\
121024A&$1.30\pm0.20$&$2.298$&$-1.36 ^{+0.16 }_{-0.16 }$&$$&$245.96 ^{+621.09 }_{-127.14 }$&$52.01 \pm0.07 $&$1,4$\\
121027A&$1.30\pm0.20$&$1.773$&$-1.49 ^{+0.43 }_{-0.39 }$&$$&$61.75 ^{+437.65 }_{-13.25 }$&$51.52 \pm0.07 $&$1,4$\\
121117A&$1.10\pm0.10$&$3.1$&$-1.30 ^{+0.11 }_{-0.11 }$&$$&$156.83 ^{+380.56 }_{-56.00 }$&$52.15 \pm0.04 $&$1,4$\\
121128A&$12.90\pm0.40$&$2.2$&$-0.80 ^{+0.07 }_{-0.07 }$&$-2.41 ^{+0.06 }_{-0.06 }$&$62.00 ^{+3.02 }_{-3.02 }$&$52.79 \pm0.01 $&$1,3$\\
121201A&$0.80\pm0.10$&$3.385$&$-1.85 ^{+0.18 }_{-0.19 }$&$$&$54.88 ^{+202.37 }_{-32.70 }$&$52.15 \pm0.05 $&$1,4$\\
121209A&$3.40\pm0.30$&$2.1$&$-1.43 ^{+0.08 }_{-0.08 }$&$$&$159.35 $&$52.24 \pm0.04 $&$1,3$\\
121211A&$1.00\pm0.30$&$1.023$&$-0.27 ^{+0.37 }_{-0.37 }$&$$&$100.00 ^{+15.00 }_{-15.00 }$&$50.80 \pm0.13 $&$1,3$\\
121229A&$0.10\pm0.00$&$2.707$&$-2.43 ^{+0.46 }_{-0.46 }$&$$&$250.00 ^{*}$&$51.53 \pm0.00 $&$2,3$\\
130131B&$1.00\pm0.20$&$2.539$&$-1.24 ^{+0.22 }_{-0.22 }$&$$&$300.42 ^{+552.02 }_{-188.93 }$&$52.06 \pm0.09 $&$1,4$\\
130215A&$2.50\pm0.70$&$0.597$&$-1.19 ^{+0.11 }_{-0.11 }$&$-1.59 ^{+0.04 }_{-0.04 }$&$257.00 ^{+130.00 }_{-130.00 }$&$51.37 \pm0.12 $&$1,3$\\
130408A&$4.90\pm1.00$&$3.757$&$-0.40 ^{+0.20 }_{-0.20 }$&$-2.30 ^{+0.20 }_{-0.20 }$&$211.00 ^{+29.00 }_{-29.00 }$&$53.33 \pm0.09 $&$1,3$\\
130418A&$0.60\pm0.20$&$1.218$&$0.27 ^{+1.88 }_{-1.42 }$&$$&$33.88 ^{+4.70 }_{-6.28 }$&$50.51 \pm0.14 $&$1,4$\\
130420A&$3.40\pm0.20$&$1.297$&$-1.52 ^{+0.15 }_{-0.15 }$&$$&$33.20 ^{+4.11 }_{-4.11 }$&$51.58 \pm0.03 $&$1,3$\\
130427A&$331.00\pm4.60$&$0.3399$&$-0.91 ^{+0.01 }_{-0.01 }$&$-3.18 ^{+0.03 }_{-0.03 }$&$877.80 ^{+4.90 }_{-4.90 }$&$53.00 \pm0.01 $&$1,3$\\
130427B&$3.00\pm0.40$&$2.78$&$-1.64 ^{+0.15 }_{-0.15 }$&$$&$102.12 $&$52.46 \pm0.06 $&$1,3$\\
130505A&$30.00\pm3.10$&$2.27$&$-0.81 ^{+0.12 }_{-0.11 }$&$$&$738.91 ^{+1073.10 }_{-401.76 }$&$53.92 \pm0.04 $&$1,4$\\
130511A&$1.30\pm0.20$&$1.3033$&$-1.54 ^{+0.33 }_{-0.34 }$&$$&$149.18 ^{+614.92 }_{-93.67 }$&$51.31 \pm0.07 $&$1,4$\\
130514A&$2.80\pm0.30$&$3.6$&$-1.44 ^{+0.17 }_{-0.15 }$&$$&$23.91 ^{+9.13 }_{-4.57 }$&$52.57 \pm0.05 $&$1,3$\\
130528A&$3.00\pm0.20$&$1.25$&$-1.08 ^{+0.05 }_{-0.05 }$&$-2.60 ^{+0.30 }_{-0.30 }$&$122.00 ^{+7.00 }_{-7.00 }$&$51.65 \pm0.03 $&$3,3$\\
130604A&$0.80\pm0.20$&$1.06$&$-1.65 ^{+0.15 }_{-0.15 }$&$$&$105.13 ^{+511.35 }_{-35.06 }$&$50.86 \pm0.11 $&$1,4$\\
130606A&$2.60\pm0.20$&$5.913$&$-1.14 ^{+0.15 }_{-0.15 }$&$$&$294.00 ^{+90.00 }_{-50.00 }$&$53.36 \pm0.03 $&$1,3$\\
130610A&$1.70\pm0.20$&$2.092$&$-1.58 ^{+0.08 }_{-0.08 }$&$$&$271.00 ^{+111.00 }_{-111.00 }$&$52.04 \pm0.05 $&$1,3$\\
130612A&$1.70\pm0.30$&$2.006$&$-1.01 ^{+2.14 }_{-2.14 }$&$-2.23 ^{+0.15 }_{-0.15 }$&$26.40 ^{+6.80 }_{-6.80 }$&$51.83 \pm0.08 $&$1,3$\\
130701A&$17.10\pm0.70$&$1.155$&$-1.10 ^{+0.06 }_{-0.06 }$&$$&$89.00 ^{+2.42 }_{-2.42 }$&$52.16 \pm0.02 $&$1,3$\\
130831A&$13.60\pm0.60$&$0.4791$&$-0.61 ^{+0.06 }_{-0.06 }$&$-2.30 ^{+0.30 }_{-0.30 }$&$55.00 ^{+4.00 }_{-4.00 }$&$51.23 \pm0.02 $&$1,3$\\
130907A&$25.60\pm0.50$&$1.238$&$-0.91 ^{+0.02 }_{-0.02 }$&$-2.34 ^{+0.07 }_{-0.07 }$&$79.50 ^{+4.96 }_{-4.96 }$&$52.55 \pm0.01 $&$1,3$\\
130925A&$7.30\pm0.60$&$0.347$&$-1.50 ^{+0.03 }_{-0.03 }$&$$&$107.00 ^{+1.81 }_{-1.81 }$&$50.60 \pm0.04 $&$2,3$\\
131030A&$28.10\pm0.70$&$1.295$&$-0.71 ^{+0.12 }_{-0.12 }$&$-2.95 ^{+0.28 }_{-0.28 }$&$177.00 ^{+10.00 }_{-10.00 }$&$52.76 \pm0.01 $&$1,3$\\
131103A&$1.50\pm0.30$&$0.599$&$-1.85 ^{+0.37 }_{-0.40 }$&$$&$40.66 ^{+224.16 }_{-39.66 }$&$50.52 \pm0.09 $&$1,4$\\
131105A&$3.50\pm0.60$&$1.686$&$-1.26 ^{+0.02 }_{-0.02 }$&$-2.33 ^{+0.33 }_{-0.33 }$&$265.00 ^{+17.00 }_{-17.00 }$&$52.23 \pm0.07 $&$1,3$\\
131117A&$0.70\pm0.10$&$4.042$&$-0.17 ^{+1.61 }_{-1.10 }$&$$&$44.40 ^{+10.40 }_{-6.00 }$&$51.90 \pm0.06 $&$1,3$\\
131227A&$1.10\pm0.20$&$5.3£¿$&$-1.39 ^{+0.13 }_{-0.13 }$&$$&$182.21 ^{+527.61 }_{-74.11 }$&$52.74 \pm0.08 $&$1,4$\\
140114A&$0.90\pm0.10$&$3$&$-2.06 ^{+0.09 }_{-0.09 }$&$$&$250.00 ^{*}$&$52.27 \pm0.05 $&$1,3$\\
140206A&$19.40\pm0.50$&$2.73$&$-0.20 ^{+0.06 }_{-0.06 }$&$-2.40 ^{+0.06 }_{-0.06 }$&$120.00 ^{+3.63 }_{-3.63 }$&$53.40 \pm0.01 $&$1,3$\\
140213A&$23.50\pm0.80$&$1.2076$&$-1.13 ^{+0.03 }_{-0.03 }$&$-2.26 ^{+0.05 }_{-0.05 }$&$86.60 ^{+3.60 }_{-3.60 }$&$52.52 \pm0.01 $&$1,3$\\
140301A&$0.70\pm0.20$&$1.416$&$-1.00 ^{*}$&$-2.14 ^{+0.36 }_{-0.42 }$&$1.01 ^{+31.81 }_{--0.02 }$&$51.33 \pm0.12 $&$1,4$\\
140304A&$1.70\pm0.20$&$5.283$&$-0.80 ^{+0.22 }_{-0.22 }$&$-2.35 ^{+0.43 }_{-0.43 }$&$123.00 ^{+27.00 }_{-27.00 }$&$52.95 \pm0.05 $&$1,3$\\
140311A&$1.30\pm0.50$&$4.954$&$-1.62 ^{+0.24 }_{-0.24 }$&$$&$77.88 ^{+250.45 }_{-27.47 }$&$52.67 \pm0.17 $&$1,4$\\
140318A&$0.50\pm0.20$&$1.02$&$-1.35 ^{+0.29 }_{-0.29 }$&$$&$128.40 ^{+434.15 }_{-60.53 }$&$50.58 \pm0.17 $&$1,4$\\
140419A&$4.90\pm0.20$&$3.956$&$-0.63 ^{+0.36 }_{-0.22 }$&$-2.30 ^{+0.40 }_{-2.50 }$&$293.00 ^{+84.00 }_{-84.00 }$&$53.45 \pm0.02 $&$1,3$\\
140423A&$2.10\pm0.20$&$3.26$&$-0.58 ^{+0.12 }_{-0.12 }$&$-1.83 ^{+0.05 }_{-0.05 }$&$121.00 ^{+15.00 }_{-15.00 }$&$52.79 \pm0.04 $&$1,3$\\
140428A&$0.60\pm0.20$&$4.7$&$-1.54 ^{+0.26 }_{-0.26 }$&$$&$250.00 ^{*}$&$52.41 \pm0.14 $&$1,3$\\
140430A&$2.50\pm0.20$&$1.6$&$-1.00 ^{*}$&$-2.03 ^{+0.25 }_{-0.27 }$&$12.25 ^{+28.12 }_{-11.25 }$&$51.90 \pm0.03 $&$1,4$\\
140506A&$10.90\pm0.90$&$0.889$&$-1.18 ^{+0.11 }_{-0.11 }$&$-2.30 ^{+0.40 }_{-7.70 }$&$197.00 ^{+32.00 }_{-32.00 }$&$52.01 \pm0.04 $&$1,3$\\
140509A&$1.60\pm0.40$&$2.4$&$-1.69 ^{+0.21 }_{-0.21 }$&$$&$74.08 ^{+242.97 }_{-25.89 }$&$52.02 \pm0.11 $&$1,4$\\
140512A&$6.80\pm0.30$&$0.725$&$-1.22 ^{+0.02 }_{-0.02 }$&$-3.20 ^{+1.60 }_{-1.60 }$&$682.00 ^{+70.00 }_{-70.00 }$&$51.82 \pm0.02 $&$1,3$\\
140515A&$0.90\pm0.10$&$6.32$&$-0.98 ^{+0.39 }_{-0.39 }$&$$&$51.30 ^{+8.89 }_{-8.89 }$&$52.56 \pm0.05 $&$1,3$\\
140518A&$1.00\pm0.10$&$4.707$&$-0.92 ^{+0.37 }_{-0.37 }$&$$&$43.90 ^{+4.60 }_{-4.60 }$&$52.28 \pm0.04 $&$1,3$\\
140629A&$4.20\pm0.40$&$2.275$&$-1.42 ^{+0.54 }_{-0.54 }$&$$&$86.01 ^{+17.53 }_{-17.53 }$&$52.32 \pm0.04 $&$1,3$\\
140703A&$2.80\pm0.60$&$3.14$&$-1.10 ^{+0.06 }_{-0.06 }$&$$&$177.00 ^{+14.01 }_{-14.01 }$&$52.60 \pm0.09 $&$1,3$\\
140710A&$1.90\pm0.30$&$0.558$&$$&$$&$$&$50.87 \pm0.07 $&$1,3$\\
140907A&$2.50\pm0.20$&$1.21$&$-0.97 ^{+0.08 }_{-0.08 }$&$$&$113.00 ^{+7.00 }_{-7.00 }$&$51.42 \pm0.03 $&$1,5$\\
141004A&$6.10\pm0.30$&$0.573$&$-1.30 ^{+0.10 }_{-0.10 }$&$$&$147.00 ^{+28.00 }_{-28.00 }$&$51.08 \pm0.02 $&$1,6$\\
141026A&$0.40\pm0.20$&$3.35$&$-1.00 ^{*}$&$-2.35 ^{+0.21 }_{-0.22 }$&$12.25 ^{+8.67 }_{-11.25 }$&$51.77 \pm0.22 $&$1,4$\\
141109A&$2.50\pm0.20$&$2.993$&$-1.56 ^{+0.07 }_{-0.07 }$&$$&$134.99 ^{+348.35 }_{-37.51 }$&$52.47 \pm0.03 $&$1,4$\\
141121A&$0.90\pm0.30$&$1.47$&$-1.79 ^{+0.15 }_{-0.15 }$&$$&$60.65 ^{+58.07 }_{-18.24 }$&$51.27 \pm0.14 $&$1,4$\\
141220A&$8.90\pm0.70$&$1.3195$&$-0.80 ^{+0.03 }_{-0.03 }$&$$&$180.00 ^{+5.44 }_{-5.44 }$&$52.21 \pm0.03 $&$1,3$\\
141221A&$3.10\pm0.20$&$1.452$&$-1.07 ^{+0.13 }_{-0.13 }$&$$&$152.40 ^{+28.50 }_{-28.50 }$&$51.79 \pm0.03 $&$1,7$\\
141225A&$1.30\pm0.40$&$0.915$&$-1.29 ^{+0.16 }_{-0.15 }$&$$&$300.42 ^{+552.02 }_{-174.05 }$&$51.08 \pm0.13 $&$1,4$\\
150120B&$1.20\pm0.20$&$3.5$&$-1.43 ^{+0.28 }_{-0.24 }$&$$&$130.00 ^{+150.00 }_{-50.00 }$&$52.29 \pm0.07 $&$1,8$\\
150206A&$10.10\pm0.40$&$2.087$&$-1.20 ^{+0.23 }_{-0.22 }$&$$&$189.51 ^{+1157.45 }_{-66.47 }$&$52.74 \pm0.02 $&$1,4$\\
150301B&$3.00\pm0.20$&$1.5169$&$-1.50 ^{+0.07 }_{-0.07 }$&$$&$149.18 ^{+369.54 }_{-47.29 }$&$51.83 \pm0.03 $&$1,4$\\
150314A&$38.50\pm0.90$&$1.758$&$-0.87 ^{+0.12 }_{-0.12 }$&$$&$358.71 ^{+381.85 }_{-108.15 }$&$53.41 \pm0.01 $&$1,4$\\
150323A&$5.40\pm0.30$&$0.593$&$-1.41 ^{+0.35 }_{-0.32 }$&$$&$81.38 ^{+176.78 }_{-17.78 }$&$50.97 \pm0.02 $&$1,4$\\
150403A&$17.60$&$2.06$&$-0.96 ^{+0.16 }_{-0.15 }$&$$&$227.86 ^{+189.81 }_{-59.89 }$&$53.05 \pm0.00 $&$1,4$\\
150413A&$1.60\pm0.30$&$3.139$&$-1.72 ^{+0.14 }_{-0.14 }$&$$&$77.88 ^{+195.76 }_{-21.93 }$&$52.32 \pm0.08 $&$1,4$\\
150727A&$1.00\pm0.20$&$0.313$&$-0.89 ^{+0.41 }_{-0.37 }$&$$&$152.40 ^{+1257.81 }_{-51.73 }$&$49.69 \pm0.09 $&$1,4$\\
150818A&$2.40\pm0.30$&$0.282$&$-1.86 ^{+0.10 }_{-0.10 }$&$$&$74.08 ^{+825.80 }_{-68.05 }$&$49.99 \pm0.05 $&$1,4$\\
150910A&$1.10\pm0.40$&$1.359$&$-1.43 ^{+0.14 }_{-0.13 }$&$$&$271.83 ^{+522.06 }_{-169.04 }$&$51.39 \pm0.16 $&$1,4$\\
150915A&$0.50\pm0.20$&$1.968$&$$&$$&$$&$51.60 \pm0.17 $&$1,3$\\
151021A&$9.70\pm0.80$&$2.33$&$-1.14 ^{+0.15 }_{-0.14 }$&$-2.46 ^{+0.18 }_{-0.32 }$&$170.00 ^{+23.00 }_{-18.00 }$&$52.91 \pm0.04 $&$1,9$\\
151027A&$6.80\pm0.60$&$0.81$&$-1.14 ^{+0.04 }_{-0.04 }$&$$&$340.00 ^{+63.00 }_{-63.00 }$&$51.73 \pm0.04 $&$1,3$\\
151029A&$1.80\pm0.30$&$1.423$&$-0.86 ^{+1.18 }_{-0.95 }$&$$&$31.34 ^{+7.13 }_{-17.25 }$&$51.26 \pm0.07 $&$1,4$\\
151031A&$1.70\pm0.20$&$1.167$&$$&$$&$$&$51.59 \pm0.05 $&$1,3$\\
151111A&$1.00\pm0.10$&$3.5$&$-0.12 ^{+5.12 }_{-0.82 }$&$-2.32 ^{+0.32 }_{-0.59 }$&$45.22 ^{+0.03 }_{-0.47 }$&$52.13 \pm0.04 $&$1,4$\\
151112A&$1.90\pm0.20$&$4.1$&$-1.87 ^{+0.17 }_{-0.17 }$&$$&$49.66 ^{+34.95 }_{-22.08 }$&$52.74 \pm0.05 $&$1,4$\\
151215A&$1.60\pm0.20$&$2.59$&$$&$$&$$&$52.39 \pm0.05 $&$1,3$\\
160121A&$1.20\pm0.20$&$1.96$&$$&$$&$$&$51.98 \pm0.07 $&$1,3$\\
160131A&$6.40\pm0.30$&$0.972$&$-1.22 ^{+0.04 }_{-0.04 }$&$$&$385.74 ^{+632.85 }_{-142.71 }$&$51.92 \pm0.02 $&$1,4$\\
160203A&$1.30\pm0.40$&$3.52$&$$&$$&$$&$52.60 \pm0.13 $&$1,3$\\
160227A&$0.60\pm0.10$&$2.38$&$$&$$&$$&$51.87 \pm0.07 $&$1,3$\\
160228A&$1.20\pm0.10$&$1.64$&$$&$$&$$&$51.79 \pm0.04 $&$1,3$\\
160314A&$0.90\pm0.20$&$0.726$&$$&$$&$$&$50.82 \pm0.10 $&$1,3$\\
160327A&$1.80\pm0.20$&$5?$&$$&$$&$$&$53.08 \pm0.05 $&$1,3$\\
160410A&$3.50\pm0.30$&$1.717$&$-0.82 ^{+0.14 }_{-0.14 }$&$$&$495.30 ^{+890.16 }_{-232.98 }$&$52.50 \pm0.04 $&$2,4$\\
160425A&$2.80\pm0.20$&$0.555$&$   $&$   $&$   $&$51.03 \pm0.03 $&$1, $\\
160804A&$2.90\pm0.30$&$0.736$&$-1.10 ^{+0.10 }_{-0.10 }$&$$&$74.00 ^{+3.00 }_{-3.00 }$&$50.87 \pm0.04 $&$1,10$\\
161014A&$2.90\pm0.60$&$2.823$&$-1.57 ^{+0.16 }_{-0.16 }$&$$&$110.52 ^{+309.25 }_{-41.46 }$&$52.45 \pm0.09 $&$1,4$\\
161017A&$2.80\pm0.20$&$2.013$&$-1.25 ^{+0.16 }_{-0.16 }$&$$&$289.00 ^{+73.00 }_{-73.00 }$&$52.25 \pm0.03 $&$1,11$\\
161108A&$0.60\pm0.10$&$1.159$&$   $&$   $&$   $&$51.13 \pm0.07 $&$1, $\\
161117A&$6.80\pm0.30$&$1.549$&$-0.43 ^{+0.07 }_{-0.07 }$&$-2.27 ^{+0.04 }_{-0.04 }$&$68.04 ^{+2.66 }_{-2.66 }$&$52.22 \pm0.02 $&$1,12$\\
161129A&$3.40\pm0.20$&$0.645$&$-1.29 ^{+0.25 }_{-0.24 }$&$$&$128.54 ^{+266.70 }_{-35.78 }$&$50.92 \pm0.03 $&$1,4$\\
161219B&$5.30\pm0.40$&$0.1475$&$-1.59 ^{+0.71 }_{-0.71 }$&$$&$91.00 ^{+21.00 }_{-21.00 }$&$49.62 \pm0.03 $&$1,13$\\
170113A&$1.10\pm0.10$&$1.968$&$-0.75 ^{+0.59 }_{-0.59 }$&$$&$73.30 ^{+33.60 }_{-33.60 }$&$51.47 \pm0.04 $&$1,14$\\
170202A&$4.70\pm0.30$&$3.645$&$-1.16 ^{+0.59 }_{-0.34 }$&$$&$247.00 ^{+166.00 }_{-86.00 }$&$53.07 \pm0.03 $&$1,15$\\
170519A&$0.70\pm0.10$&$0.818$&$   $&$   $&$   $&$50.83 \pm0.06 $&$1, $\\
170531B&$0.80\pm0.20$&$2.366$&$   $&$   $&$   $&$51.99 \pm0.11 $&$1, $\\
170604A&$4.20\pm0.90$&$1.329$&$-1.40 ^{+0.13 }_{-0.12 }$&$$&$245.96 ^{+621.09 }_{-115.66 }$&$51.93 \pm0.09 $&$1,4$\\
170705A&$13.90\pm0.40$&$2.01$&$-0.88 ^{+0.07 }_{-0.07 }$&$-2.38 ^{+0.09 }_{-0.09 }$&$100.00 ^{+10.00 }_{-10.00 }$&$52.83 \pm0.01 $&$1,16$\\
170903A&$3.90\pm0.50$&$0.886$&$$&$$&$$&$51.66 \pm0.06 $&$1,$\\
171020A&$0.70\pm0.20$&$1.87$&$   $&$   $&$   $&$51.70 \pm0.12 $&$1, $\\
171222A&$0.70\pm0.20$&$2.409$&$-1.09 ^{+0.27 }_{-0.27 }$&$$&$29.18 ^{+2.73 }_{-2.73 }$&$51.45 \pm0.12 $&$1,17$\\
180115A&$0.60\pm0.20$&$2.487?$&$   $&$   $&$   $&$51.92 \pm0.14 $&$1, $\\
180205A&$3.40\pm0.30$&$1.409$&$-1.38 ^{+0.15 }_{-0.15 }$&$$&$85.00 ^{+14.40 }_{-14.40 }$&$51.70 \pm0.04 $&$1,18$\\
180314A&$7.90\pm0.60$&$1.445$&$-0.61 ^{+0.17 }_{-0.16 }$&$$&$111.00 ^{+6.00 }_{-6.00 }$&$52.10 \pm0.03 $&$2,19$\\
180325A&$9.40\pm0.30$&$2.248$&$-0.50 ^{+0.21 }_{-0.19 }$&$-2.65 ^{+0.31 }_{-1.06 }$&$306.00 ^{+50.00 }_{-39.00 }$&$53.18 \pm0.01 $&$1,20$\\
180329B&$1.40\pm0.40$&$1.998$&$-0.97 ^{+0.56 }_{-0.56 }$&$$&$48.60 ^{+9.10 }_{-9.10 }$&$51.55 \pm0.12 $&$1,21$\\
180404A&$1.30\pm0.20$&$1$&$   $&$   $&$   $&$51.31 \pm0.07 $&$1, $\\
180510B&$1.20\pm0.20$&$1.305$&$   $&$   $&$   $&$51.56 \pm0.07 $&$1, $\\
180620B&$3.60\pm0.20$&$1.1175$&$-0.50 ^{+0.20 }_{-0.20 }$&$-2.20 ^{+0.10 }_{-0.10 }$&$106.00 ^{+15.00 }_{-15.00 }$&$51.74 \pm0.02 $&$1,22$\\
180624A&$1.40\pm0.20$&$2.855$&$-1.89 ^{+0.10 }_{-0.10 }$&$$&$54.88 ^{+360.57 }_{-47.74 }$&$52.24 \pm0.06 $&$1,4$\\
180728A&$132.00\pm2.90$&$0.117$&$-1.54 ^{+0.01 }_{-0.01 }$&$-2.46 ^{+0.02 }_{-0.02 }$&$79.20 ^{+1.40 }_{-1.40 }$&$50.86 \pm0.01 $&$1,23$\\
181010A&$1.40\pm0.20$&$1.39$&$-0.80 ^{+0.20 }_{-0.20 }$&$$&$280.00 ^{+80.00 }_{-80.00 }$&$51.64 \pm0.06 $&$1,24$\\
181020A&$7.70\pm0.30$&$2.938$&$-0.73 ^{+0.13 }_{-0.11 }$&$-2.35 ^{+0.22 }_{-0.41 }$&$371.00 ^{+57.00 }_{-52.00 }$&$53.41 \pm0.02 $&$1,25$\\
181110A&$3.70\pm0.30$&$1.505$&$-1.63 ^{+0.34 }_{-0.26 }$&$$&$48.00 ^{+14.00 }_{-27.00 }$&$51.83 \pm0.04 $&$1,26$\\
190106A&$5.50\pm0.30$&$1.859$&$-1.00 ^{+0.63 }_{-0.50 }$&$$&$171.00 ^{+90.00 }_{-42.00 }$&$52.34 \pm0.02 $&$1,27$\\
190114A&$0.50\pm0.20$&$3.3765$&$$&$$&$$&$52.15 \pm0.17 $&$1,$\\
190114C&$101.00\pm1.50$&$0.425$&$-1.06 ^{+0.00 }_{-0.00 }$&$-3.18 ^{+0.07 }_{-0.07 }$&$998.60 ^{+11.90 }_{-11.90 }$&$52.66 \pm0.01 $&$1,28$\\
190324A&$11.90\pm0.50$&$1.1715$&$-0.84 ^{+0.04 }_{-0.04 }$&$-2.06 ^{+0.05 }_{-0.05 }$&$144.30 ^{+8.40 }_{-8.40 }$&$52.41 \pm0.02 $&$2,29$\\
190613A&$1.50\pm0.70$&$2.78?$&$0.06 ^{+0.18 }_{-0.18 }$&$$&$108.00 ^{+5.00 }_{-5.00 }$&$52.08 \pm0.20 $&$1,30$\\
190719C&$5.50\pm0.30$&$2.469$&$-0.87 ^{+0.18 }_{-0.18 }$&$$&$81.00 ^{+9.00 }_{-9.00 }$&$52.44 \pm0.02 $&$1,31$\\
190829A&$18.00\pm2.70$&$0.0785$&$-1.33 ^{+0.30 }_{-0.23 }$&$$&$579.00 ^{+2282.00 }_{-281.00 }$&$49.94 \pm0.07 $&$1,32$\\
191004B&$5.00\pm0.20$&$3.503$&$-0.51 ^{+0.34 }_{-0.30 }$&$$&$172.00 ^{+27.00 }_{-21.00 }$&$53.01 \pm0.02 $&$2,33$\\
191011A&$1.80\pm0.20$&$1.722$&$-1.24 ^{+0.29 }_{-0.29 }$&$$&$89.00 ^{+23.00 }_{-23.00 }$&$51.62 \pm0.05 $&$2,34$\\
191019A&$5.70\pm0.40$&$0.248$&$   $&$   $&$   $&$50.53 \pm0.03 $&$2, $\\

\hline
\end{longtable}

\begin{description}
\item Note. Columns show, in order, the GRB name; the peak photon flux in the 15--150 keV {\it Swift}/BAT energy band
(with 90\% confidence level error); the redshift; various spectral parameters (with 68\% confidence level errors), including the low-energy photon index $\alpha$, the high-energy photon index $\beta$, and the observer-frame peak energy $E_p$; and the peak bolometric luminosity in units of erg s$^{-1}$ (calculated in the 1--$10^4$ rest-frame energy range, with 90\% confidence level error that propagated from the error of the peak flux only). $^{(*)}$ Bursts with missing $\alpha$ or $E_p$ are reported with the typical values of $-1$ or $250$ keV. In the last column, we give
     the references, in order, for the redshift and for the spectral parameters: 1) \href{https://www.mpe.mpg.de/~jcg/grbgen.html}{https://www.mpe.mpg.de/~jcg/grbgen.html}; 2) \href{https://swift.gsfc.nasa.gov/archive/grb_table/}{https://swift.gsfc.nasa.gov/archive/grb$\underline{ }$table/}; 3) \cite{2020ApJ...893...77W}; 4) \cite{2007ApJ...671..656B}; 5) \cite{2014GCN.16798....1Z}; 6) \cite{2014GCN.16900....1P}; 7) \cite{2014GCN.17216....1Y}; 8) \cite{2015GCN.17319....1V}; 9) \cite{2015GCN.18433....1G}; 10) \cite{2016GCN.19769....1B}; 11) \cite{2016GCN.20082....1F}; 12) \cite{2016GCN.20192....1M}; 13) \cite{2016GCN.20323....1F}; 14) \cite{2017GCN.20456....1M}; 15) \cite{2017GCN.20604....1F}; 16) \cite{2017GCN.21297....1B}; 17) \cite{2017GCN.22277....1S}; 18) \cite{2018GCN.22386....1V}; 19) \cite{2018GCN.22513....1T}; 20) \cite{2018GCN.22546....1F}; 21) \cite{2018GCN.22566....1P}; 22) \cite{2018GCN.22813....1P}; 23) \cite{2018GCN.23053....1P}; 24) \cite{2018GCN.23320....1V}; 25) \cite{2018GCN.23363....1T}; 26) \cite{2018GCN.23424....1F}; 27) \cite{2019GCN.23637....1T}; 28) \cite{2019GCN.23707....1H}; 29) \cite{2019GCN.24002....1H}; 30) \cite{2019GCN.24816....1P}; 31) \cite{2019GCN.25130....1P}; 32) \cite{2019GCN.25660....1T}; 33) \cite{2019GCN.25974....1S}; 34) \cite{2019GCN.26000....1B}.
\end{description}



\bsp	
\label{lastpage}
\end{document}